\documentclass[usletter,11pt]{article}

\usepackage{amssymb}
\usepackage{amsmath}
\usepackage{graphicx}
\usepackage{subfigure}
\usepackage{psfrag}
\usepackage[round]{natbib}
\usepackage[colorlinks]{hyperref}
\hypersetup{colorlinks,breaklinks=true,citecolor=blue,filecolor=blue,linkcolor=blue,urlcolor=blue,pdfauthor={C.~Guervilly}}
\usepackage{appendix}

\newcommand{\vel}{\mathbf{u}}
\newcommand{\B}{\mathbf{B}}
\newcommand{\pdt}[1]{\frac{\partial #1}{\partial t}}
\newcommand{\n}{\nabla}

\newcommand{\pl}{\left(}
\newcommand{\pr}{\right)}
\newcommand{\vect}[1]{\mathbf{#1}}
\newcommand{\ie}{\textit{i.e.}}
\newcommand{\eg}{\textit{e.g.}}

\begin{document}

\title{A dynamo driven by zonal jets at the upper surface: Applications to giant planets}
\author{C\'eline {Guervilly}$^{a,b}$,Philippe {Cardin}$^{a}$, Nathana\"el {Schaeffer}$^{a}$
	\\ {\small $^{a}$ ISTerre, Universit\'e de Grenoble 1/CNRS, F-38041, Grenoble, France}
	\\ {\small $^{b}$ Department of Applied Mathematics and Statistics, Baskin School of Engineering,}
	\\ {\small University of California, Santa Cruz, CA 95064, USA}
	}
\date{January 18, 2012}

\maketitle

\begin{abstract}
We present a dynamo mechanism arising from the presence of barotropically unstable zonal jet currents
in a rotating spherical shell. 
The shear instability of the zonal flow develops in the form of a global Rossby mode, whose azimuthal wavenumber depends on the width of the zonal jets.
We obtain self-sustained magnetic fields at magnetic Reynolds numbers greater than $10^3$.
We show that the propagation of the Rossby waves is crucial for dynamo action.
The amplitude of the axisymmetric poloidal magnetic field depends on the wavenumber of the Rossby mode, and hence on the width of the zonal jets.
We discuss the plausibility of this dynamo mechanism for generating the magnetic field of the giant planets.
Our results suggest a possible link between the topology of the magnetic field and the profile of the zonal winds observed at the surface of the giant planets.
For narrow Jupiter-like jets, the poloidal magnetic field is dominated by an axial dipole whereas for wide Neptune-like jets, the axisymmetric poloidal field is weak.
\end{abstract}

\section{Introduction}

\begin{figure}
 \centering
    \includegraphics[clip=true,width=0.8\textwidth]{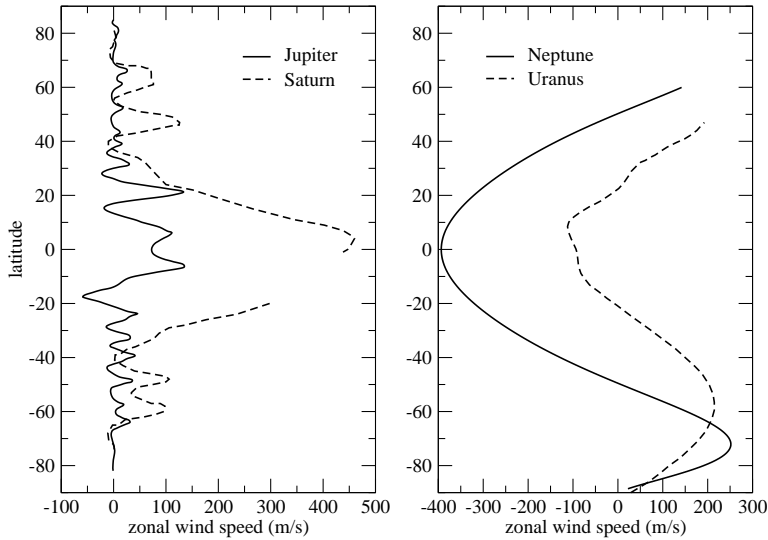}
    \caption{Zonal velocity measured at the surface in the planet's mean rotating frame for each of the four giants
by tracking cloud features in the outer weather layer. Profiles adapted from \citet{Por03}, \citet{San00}, \citet{Sro01} and \citet{Sro05}.}
 \label{fig:prof_surf}
\end{figure}

The zonal (\ie\ axisymmetric and azimuthally directed) jet streams visible at the surface of the giant planets are a persistent feature
of the fluid dynamics of these planets (figure~\ref{fig:prof_surf}). 
The gas giants (Jupiter and Saturn) display a strong eastward equatorial jet, extending 
to latitudes $\pm 20^\circ$ with a peak velocity exceeding $100$ m/s on Jupiter \citep{Por03}, 
and to latitudes $\pm 30^\circ$ with a peak velocity exceeding $400$ m/s on Saturn \citep{San00}. 
At higher latitudes, alternating prograde (eastward) and retrograde 
(westward) jets of smaller amplitude are observed extending all the way to the poles. 
These profiles are fairly symmetric with respect to the equator. 
On the ice giants (Uranus and Neptune) the picture is rather different. A very intense retrograde 
equatorial current is present with maximum velocity of $100$ m/s on Uranus \citep{Sro05} and $400$ m/s on Neptune \citep{Sro01}.
At higher latitudes, a single prograde jet of large
amplitude is present in each hemisphere.
Several decades of observations show that these zonal flows remain approximately steady \citep{Por03}.

The origin of these zonal flows and the associated question of the depth to which they extend into
the planets' interiors have been areas of active research in rotating fluid dynamics
for several decades (\eg\ \citealp{Jon09}, and references therein; see also the review by \citealp{Vas05}).
In particular, several models have been proposed to explain the zonal wind pattern of Jupiter, and 
can be categorized into two main classes: weather layer models and deep convective layer models.
The former assume that the zonal flows are produced in a shallow stably stratified region near cloud level.
These models are able to reproduce the high latitude structures with alternating eastward and westward jets
and a strong equatorial current \citep[\eg][]{Wil78,Cho96}.
These models tend to produce a retrograde equatorial jet \citep{Yan03}, so they provide a plausible explanation
for the retrograde equatorial flow of the ice giants but not for the prograde flow observed on gas giants.
A parametrized forcing such as a strong equatorially-localized baroclinicity is required
to force a shallow system to produce a prograde equatorial jet \citep{Wil03}.
The second class of models is deep convection models which simulate most or all of the whole $10^4$km-thick 
molecular hydrogen layer \citep{Bus76,Chr01b,Chr02,Man96}. 
The presence of deep convection is inferred from the observation that the atmospheres of the major planets emit more energy 
by long-wave radiation than they absorb from the Sun. 
Consequently their atmospheres must receive additional heat supplied by the interior of the planet.
Recent numerical models using either a Boussinesq approximation \citep{Hei05} or 
an anelastic approximation \citep{Jon09} and low Ekman numbers (i.e. strong rotational effect
compared with viscous dissipation) display alternating zonal jets at high latitudes. 
A strong eastward equatorial jet is a robust feature of these models where
the Coriolis force dominates buoyancy, in good agreement with the gas giant observations. 
Interestingly, deep convection models suggest that the zonal velocity generated 
by non-linear interactions of convective motions (\ie\ the motions directly 
forced by buoyancy)
is roughly geostrophic, that is, invariant along the direction of the rotation axis.
This feature is also present in strongly compressible models 
provided that the Ekman number is small enough,
despite the increase of density with depth yielding ageostrophic convective motions
\citep{Jon09,Kas09}. 
When the convection is more vigorous such that the buoyancy force overcomes the Coriolis force, 
3D turbulence homogenizes angular momentum; a retrograde 
jet forms in the equatorial region and a single strong prograde jet forms in the polar region, in good agreement with the 
ice giant observations \citep{Aur07}.

\begin{figure}
 \centering
   \includegraphics[clip=true,width=0.8\textwidth]{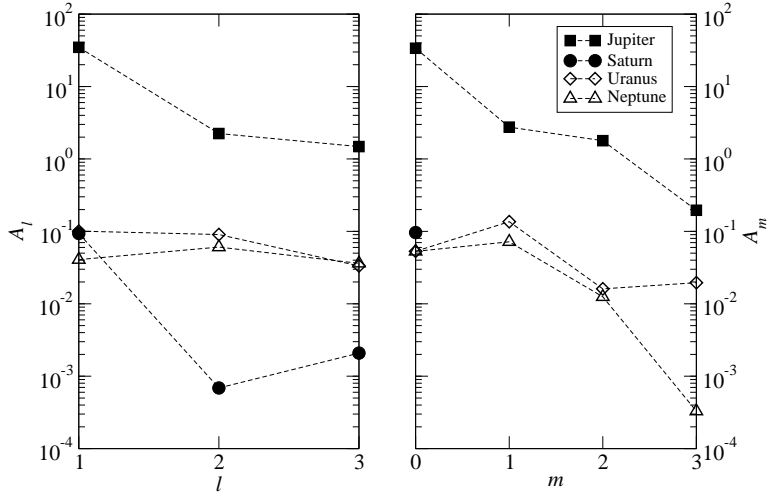}
    \caption{Spectra of the magnetic field squared amplitude at the planetary radius for degrees $l$ and order $m$ up to $3$ obtained from inversion models of the magnetic measurements. 
The squared amplitude for a given degree $l$ is \mbox{$A_l=\sum_{m=0}^l (l+1) \left[ (g_l^m)^2 + (h_l^m)^2 \right]$} using a Schmidt normalisation for the spherical harmonics.
The squared amplitude for a given mode $m$ is \mbox{$A_m=\sum_{l=m}^{l_{max}} (l+1) \left[ (g_l^m)^2 + (h_l^m)^2 \right]$}. 
$g_l^m$ and $h_l^m$ are the Gauss coefficients in gauss. After \citet{Yu10} (Model Galileo 15), \citet{Bur09} (Cassini measurements), \citet{Con91} (model O$_8$) and
\citet{Her09} (AH$_5$ model from magnetic observations and auroral data).}
\label{fig:mag_spec}
\end{figure}

Another feature of the giant planets is their strong magnetic fields (figure~\ref{fig:mag_spec}).
The observed magnetic fields for gas and ice giants differ drastically 
\citep[see for instance the recent review by][]{Rus10}. 
Jupiter and Saturn have a main axial dipole component (corresponding to $l=1$, $m=0$
in figure~\ref{fig:mag_spec}), a feature shared with the Earth for instance \citep{Yu10,Bur09}.
Neptune and Uranus, on the other hand, have strong non-axial multipolar components (corresponding to $l=2,3$
in figure~\ref{fig:mag_spec}) compared with the axial dipole component \citep{Con91,Her09}. 
The magnetic field is generated in the deep, electrically conducting regions of the planets' interiors: 
a metallic hydrogen layer for Jupiter and Saturn \citep[][and references therein]{Nel99,Gui05} and an 
electrolyte layer composed of water, methane and ammonia \citep{Hub91,Nel97} or superionic water \citep{Red11}
for Uranus and Neptune.

Numerical models of convective dynamos in rapidly rotating spherical shells typically produce 
axial dipolar dominated magnetic fields 
for moderate Rayleigh numbers and moderate Ekman numbers \citep[\eg][]{Ols99,Aub04,Chr07}. 
To explain the unusual large scale non-dipolar magnetic fields of Uranus and Neptune, models using peculiar parameter regimes or different convective region geometries have been proposed. The latter models show that a numerical dynamo operating in a thin shell surrounding a stably-stratified fluid interior produces magnetic field morphologies similar to those of Uranus and Neptune \citep{Hub95,Holme96,Sta06}.
\citet{Gom07} obtain weakly dipolar and strongly tilted dynamo magnetic fields when high magnetic diffusivities are used (or equivalently small electrical conductivity). Their results show that these peculiar fields are stable in the presence of strong zonal circulation and when the flow has a dominant effect over the magnetic fields. This feature is also emphasized by \citet{Aub04} who find stable equatorial dipole solutions with a weak magnetic field strength and low Elsasser number (measure of the relative importance of the Lorentz and Coriolis forces) for moderately low Ekman numbers.
They argue that the magnetic field geometry of the equatorial dipole solution is incompatible with the columnar convective motions and thus this morphology is stable only when Lorentz forces are weak. 

Although scaling laws derived from numerical simulations of dynamos driven by basal heating convection
predict dipolar magnetic field in planetary parameter regimes \citep{Ols06}, recent numerical simulations using 
more realistic parameter values (lower Ekman numbers) have not produced large scale magnetic fields so far,
and require larger magnetic Reynolds numbers (measure of magnetic induction versus magnetic diffusion) \citep{Kag08}.
Moreover, convection in the interior of Jupiter is often thought to be driven by secular cooling \citep{Ste03}.
Numerical dynamos driven by secular cooling typically produce weak dipole or multipolar magnetic field for larger forcing
\citep{Kut00,Ols06} depending on boundary conditions \citep{Hor10}.
Therefore the question of the generation of large scale magnetic field by turbulent convective motions in the planetary parameter regime remains open.

The dichotomies observed in the magnetic fields and in the zonal wind profiles of the giant planets are
rather striking. Up to now no study has tried to relate them directly, probably because 
the former is a feature of the deep interior whereas the latter is a characteristic 
of the surface.
However, if some mechanism is able to transport angular momentum from the surface down to the deep, fully conducting region 
then the zonal motions may influence the generation of the magnetic field.
In the non-magnetic deep convection models~\citep{Hei05,Jon09}, zonal motions extend
geostrophically throughout the electrically insulating molecular hydrogen layer down to the bottom of the model.
On the other hand, due to the possible rapid increase of electrical conductivity with depth in the outer region, 
\citet{Liu08} argued that the ohmic dissipation produced by geostrophic zonal motions shearing
dipolar magnetic field lines would exceed the luminosity measured at the surface of Jupiter if the vertical extent of this
geostrophic zonal motions exceeds 4\% of the planet radius. 
However, the argument of \citet{Liu08} is purely kinematic, that is 
the action of the magnetic forces on the flow and the feedback on the magnetic field are ignored.
In a self-consistent
magnetohydrodynamic model, the zonal flow would adjust toward a non-geostrophic state due to the action of magnetic forces if 
the electrical conductivity of the fluid is significant (\citet{Gla08}, see also 
the non-linear numerical simulations of convectively-driven dynamos of \citet{Aub05}). 
In this case,
angular momentum may be transported along the magnetic field lines leading to a dynamical state close to the 
Ferraro state. This state minimizes the ohmic dissipation produced by the shearing 
of the poloidal magnetic field by the zonal flow
as the poloidal magnetic field lines are aligned with angular velocity contours. 
Both scenarios, either geostrophic zonal balance or Ferraro state, imply the existence of multiple
zonal jets of significant amplitude at the top of the fully conducting region beneath.
The plausibility of each scenario depends on the radial profile of electrical conductivity, which is currently not
well constrained within the giant planets \citep{Nel99}.

The idea of the work presented in this paper is that these zonal jets may exert,
by viscous or electromagnetic coupling, an external forcing at the top of the
deeper conducting envelope.
From previous studies \citep{Sch06,Gue10} we know that
the viscous coupling between a differentially rotating boundary and a low-viscosity electrically conducting fluid 
can generate a self-sustained magnetic field in different geometries.
Zonal motions can be subject to barotropic shear instabilities which have a lengthscale 
independent of the viscosity, unlike convective instabilities.  
These instabilities are able to generate large scale magnetic fields, and so they are an interesting source of dynamo action 
under planetary interior conditions.
In order to test the plausibility of a dynamo driven by this source in isolation, 
we use an incompressible 3D numerical dynamo model with a zonal velocity profile imposed
at the top of a spherical shell containing a conducting fluid. 
We use a dynamical approach, that is non-linear interactions between the flow and the magnetic field are taken into account; 
therefore the fluid flow is free to adopt a three-dimensional structure as long as it satisfies the imposed viscous boundary conditions.

The dynamics of the deep conducting region is usually assumed to be slower than the dynamics of the outer molecular
 hydrogen region due to magnetic braking, even if uncertainties remain in the electrical conductivity.
The model presented in this paper assumes an idealized one-way coupling between the outer and deep regions. A more realistic model would 
need to account for the back reaction of the deep layer onto the outer layer; a study of the consistent dynamical interaction of the two layers 
is beyond the scope of this paper.
For studies of more realistic coupling, see 
promising recent numerical models of self-consistent convectively-driven dynamos in spherical shells including radially variable electrical
conductivity of \citet{Hei11} and \citet{Sta10}.
In these models, slow convective motions in the interior dynamo region coexist with strong zonal flow near the outer surface.
Differential rotation in the interior is only partially inhibited by the strong magnetic field.

In order to assess the role of the zonal wind profile on the topology of the sustained 
magnetic field, we use both Jupiter-like and Neptune-like zonal wind profiles. 
In the giant planets, as in rocky planets, it is usually assumed that the dynamo mechanism
is driven by convective motions. 
The giant planets display a strong surface heat flux (with the exception of Uranus)
meaning that heat transfer is efficient in the interior of the planet and thus mostly due to convection \citep[][and references therein]{Gui07}.
Here we want to assess the efficiency of zonal velocity forcing alone, so we do not model 
convective motions.

The first goal of this work is to quantify what 
amplitude of the zonal wind \textit{inside} the conducting layer is needed to trigger the dynamo instability,
so we do not model the exact or realistic coupling between the molecular hydrogen upper layer and the deep, electrically conducting region.
Our second goal is to test to what extent the pattern of the zonal 
flow imposed at the top of the conducting layer influences the topology of the self-sustained
magnetic field.

We first describe the model and the numerical method used (section~\ref{sec:model_jupiter}). 
Then we present numerical results from simulations in the non-magnetic case (section~\ref{sec:hydro_jupiter}) 
followed by results from dynamo simulations (section~\ref{sec:dynamo_jupiter}).
The application of our results to planetary conditions is discussed in section~\ref{sec:Summary}.

\section{Model}
\label{sec:model_jupiter}
We model the deep conducting layer of the giant planets as a thick spherical shell. 
At the top of the conducting layer we impose an axisymmetric azimuthal velocity
to represent the zonal flow generated in the overlying envelope. 
The shell rotates around the $z$-axis at the imposed rotation rate $\Omega$. 
The aspect ratio is $\gamma=r_i/r_o$ where $r_i$ is the inner sphere radius, corresponding to a
rocky core, and $r_o$ the outer sphere radius, corresponding to the top of the fully conducting region.
The fluid is assumed incompressible with constant density $\rho$ and constant temperature, that is, 
no convective motions are computed. 
The assumption of incompressibility is made for simplicity, although
the pressure scale height at the depths of the conducting layer is roughly $8000$km \citep{Gui04}, that is, about $1/5$ of the thickness of the layer.
The effects of compressibility may well play a role in the dynamics of the conducting regions \citep[see for instance][]{Evo04}.

For simplicity we model the angular momentum coupling with the external zonal flow as a  
rigid boundary condition for the velocity at the outer boundary, rather than as a shear stress condition.
The flow is driven through a boundary forcing rather than a volume forcing to avoid directly 
imposing bidimensionality to the velocity field.
As we are interested in the bulk magnetohydrodynamical
 process, the exact nature of the coupling 
(electromagnetic or viscous, shear stress or rigid) with the upper molecular hydrogen layer is not crucial for our study.
We discuss the implication of the choice of the rigid boundary condition in section~\ref{sec:hydro_jupiter}.
The radial profile of electrical conductivity is not well constrained in the gas giants. In particular 
the existence of a first order or continuous transition between the molecular and metallic hydrogen phase 
is still an open question, although
high-pressure experiments are in favor of a continuous transition \citep{Nel99}.
We choose to model the outer boundary as electrically insulating to simplify the coupling between the layers.
The conductivity is assumed constant throughout the whole modeled conducting layer.
As we do not model the molecular hydrogen layer, we assume 
zonal geostrophic balance within this envelope for simplicity. The amplitude of the zonal motions at the outer boundary of our model
is therefore the same as the surface winds. 
This idealized representation of the dynamics of the molecular hydrogen layer would be altered if the magnetic forces upset the zonal 
geostrophic balance. Depending on the magnitude and radial profile of the electrical conductivity, the amplitude of the zonal motions 
might be reduced, and the zonal flow contours would tend to align with the magnetic field lines, although we do not expect the characteristics 
of the zonal jets (narrow or wide, relative amplitude of the peaks) to be altered very much. 

We use two different synthetic azimuthal velocity profiles for the boundary forcing imposed at the top: 
a multiple jet profile for the gas giants with a 
profile based on Jupiter's surface zonal winds (hereafter profile J) and a 3-band profile 
based on Neptune's surface zonal winds (profile N).

\begin{figure}
 \centering
  \includegraphics[clip=true,width=0.8\textwidth]{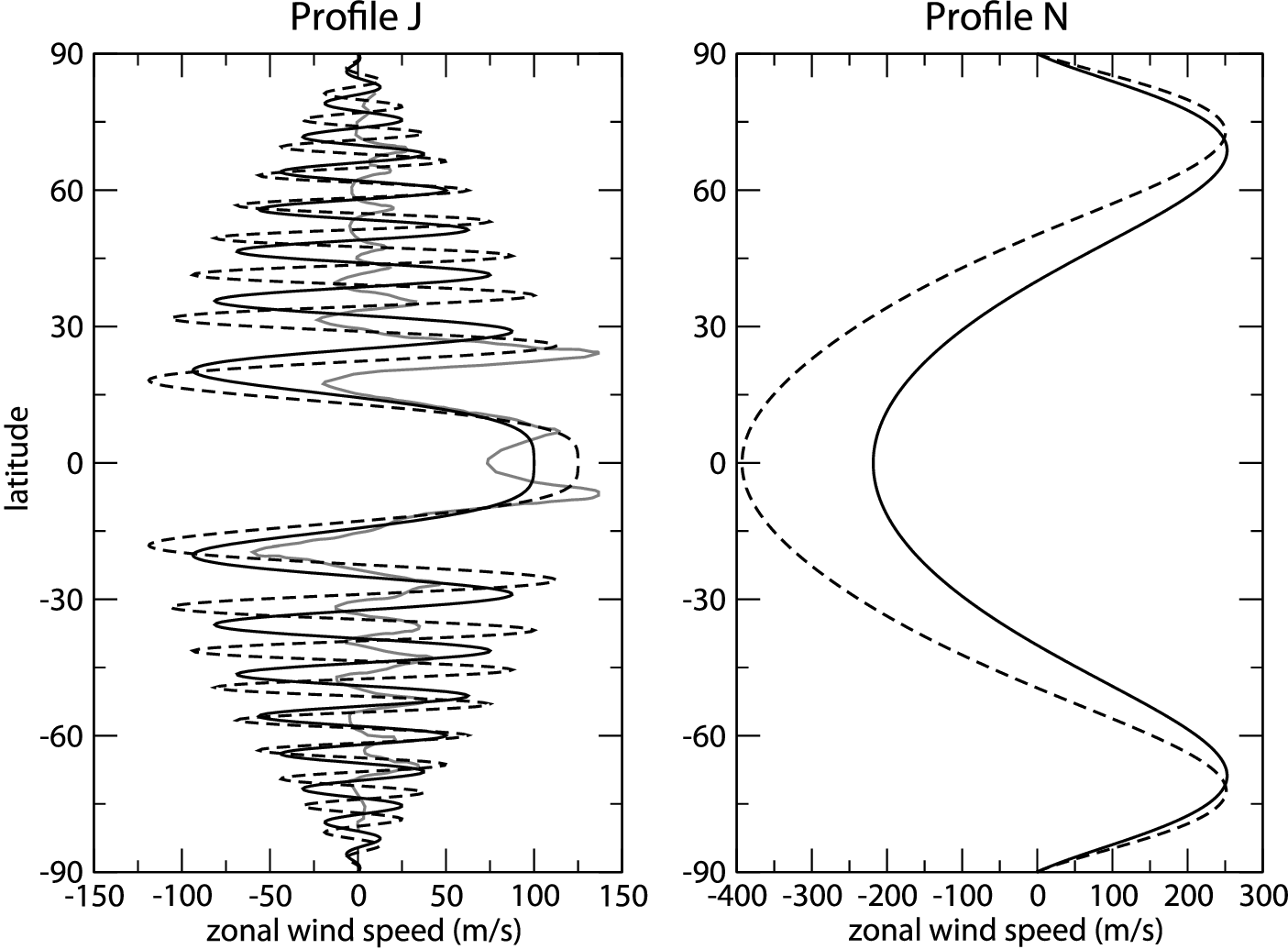}
  \caption{Zonal velocity profile imposed at the surface of model J (left) and model N (right) (solid lines).
  Both profiles are obtained by assuming that the zonal velocities are geostrophic for $r_s>r>r_o$ and using the profile
  represented by a dashed line at the surface of the planet ($r=r_s$): model J, profile (\ref{eq:profileJ}) with $n_0=4$, $\gamma_s=r_s/r_o=1.25$ and $U_0=100$;
  model N: polynomial fit of order $10$ in latitude of the zonal wind profile measured at the surface of Neptune (figure~\ref{fig:prof_surf}) with $\gamma_s=1/0.85=1.18$. 
 For comparison the zonal wind profile measured at the surface of Jupiter is plotted in gray.}
 \label{fig:prof_model}
\end{figure}

For Jupiter, we use the profile given in \citet{Wic02b}
\begin{eqnarray}
 \vect{U}=U(s)\vect{e_{\phi}}= U_0 \frac{s}{r_0 \cos(n_0\pi)} \cos\pl n_0 \pi \frac{s-r_0}{r_s-r_0} \pr \vect{e_{\phi}},
\label{eq:profileJ}
\end{eqnarray}
where $s=r\sin\theta$, $r_s$ is the surface radius of the planet and \mbox{$U_0=U(r_0,\theta=\pi/2)$}. $n_0$ controls the numbers of jets. 
The profile at the radius $r_s$ best matches the observed profile at the 
surface for $n_0=4$ (figure~\ref{fig:prof_model}). 
The profile $U(r_o,\theta)$ is used to drive the flow at the top of our simulated metallic hydrogen layer 
(figure~\ref{fig:prof_model}). 
The ratio $\gamma_s=r_s/r_o$ determines the $U$ profile at $r_o$.
We choose $\gamma_s=r_s/r_o=1/0.8=1.25$ following \citet{Gui94}.

For the Neptune-like profile, we use the zonal velocity profile measured at the surface of Neptune, 
approximated by a polynomial of order $10$ in latitude.
We project this surface velocity profile geostrophically down to $r_o$ using $\gamma_s=1/0.85=1.18$ \citep{Hub91} (figure~\ref{fig:prof_model}). 

The existence of a rocky core at the centre of the giant planets is uncertain and depends 
on the poorly constrained composition of the planet. Estimates for the core mass are
 $0-14 m_{\oplus}$ for Jupiter (total mass $318 m_{\oplus}$), 
$6-17 m_{\oplus}$ for Saturn (total mass $95 m_{\oplus}$) 
and $0-4 m_{\oplus}$ for Uranus and Neptune (total mass $15 m_{\oplus}$ and $17 m_{\oplus}$ respectively)
where $m_{\oplus}$ denotes the mass of the Earth \citep{Gui05}.
If present, the rocky cores are therefore believed to be small.
Following the interior model of Jupiter proposed by \citet{Gui94} we use an aspect ratio $r_i/r_o=0.2$ for all
the simulations performed. 
The inner core is assumed to be electrically conducting, with the same conductivity as the fluid in the conducting layer.
We did not carry out simulations with an insulating core as the effect of the conductivity of the inner 
core on the dynamo mechanism is believed to be small \citep{Wic02}.
The velocity boundary condition is no-slip at the inner boundary.

The velocity $\vel$ is scaled by $U_0$, the absolute value of the azimuthal velocity imposed at the equator 
of the outer sphere. The lengthscale is the radius of the outer sphere $r_o$. 
The magnetic field $\B$ is scaled by $\sqrt{\rho \mu_0 r_o \Omega U_0}$ where $\rho$ is the fluid density 
and $\mu_0$ is the vacuum magnetic permeability. 
\\We numerically solve the momentum equation for an incompressible fluid,
\begin{eqnarray}
Re \pdt{\vel}+ Re \pl \vel\cdot\boldsymbol{\n} \pr \vel 
+ \frac{2}{E} \mathbf{e_z}\times \vel 
= -\boldsymbol{\n}p +\boldsymbol{\n^2}\vel
+ \frac{1}{E} \pl \boldsymbol{\n} \times \B \pr \times \B ,
\label{eq:NS_Jupit}
\end{eqnarray}
the continuity equation,
\begin{eqnarray}
\boldsymbol{\n} \cdot \vel = 0 ,
\label{eq:divu_Jupit}
\end{eqnarray}
and the magnetic induction equation,
\begin{eqnarray}
\pdt \B = \boldsymbol{\n} \times \pl \vel \times \B \pr 
+ \frac{1}{Re Pm} \boldsymbol{\n^2}\B ,
\label{eq:ind_Jupit}
\end{eqnarray}
\begin{eqnarray}
 \boldsymbol{\n} \cdot \B =0 ,
\end{eqnarray}
where $p$ is the dimensionless pressure, which includes the centrifugal potential.
\newline \indent
The Reynolds number $Re=r_o U_0 / \nu$ parametrizes the mechanical forcing exerted 
on the system by controlling the amplitude of the zonal velocity.
The magnetic Prandtl number $Pm=\nu / \eta$ measures the ratio of viscous to magnetic diffusivities. 
The magnetic Reynolds number $Rm$ is defined as $Rm=Re Pm$.
The Ekman number $E=\nu / (\Omega r_o^2)$ measures the importance of the viscous term over 
the Coriolis force.
The Rossby number $Ro=Re E=U_0/(\Omega r_o)$ is the ratio of inertial force to Coriolis force.
Note that in our definition the Rossby number refers to the amplitude of the prescribed zonal jets at the surface, 
and not to the local flow velocity.
\newline
The results presented in this paper were obtained with the PARODY code, a fully three-dimensional and non-linear code. 
The code was derived from \citet{Dor97} by J. Aubert, P. Cardin, E. Dormy in the dynamo benchmark \citep{Chr01},
 and parallelised and optimised by J. Aubert and E. Dormy. The velocity and magnetic
 fields are decomposed into poloidal and toroidal scalars and expanded in spherical
 harmonic functions in the angular coordinates with $l$ representing the latitudinal degree and $m$ the azimuthal order. A finite difference scheme is used on an
 irregular radial grid (finer near the boundaries to resolve the boundary layers). A
 Crank-Nicolson scheme is implemented for the time integration of the diffusion terms and
 an Adams-Bashforth procedure is used for the other terms.

\section{Dynamics without the magnetic field}
\label{sec:hydro_jupiter}

For a rapidly rotating system in which the Coriolis force exactly balances the pressure force, the Proudman-Taylor constraint states that the flow 
is $z$-invariant and follows geostrophic contours. For an incompressible fluid
in a bounded container, these geostrophic contours correspond to surfaces of equal height. 
In a sphere the only geostrophic motions are azimuthal and axisymmetric. 
In the giant planets' conducting envelopes, the Ekman number is about $10^{-16}$ and 
the Rossby number is much smaller than $1$ \citep{Gui04}. In the absence of a magnetic field,
we expect the Proudman-Taylor constraint to hold for large scale motions. 
As we want to reach the dynamical regime in which the flow is 
strongly geostrophic, the use of small Ekman and Rossby numbers is required. 
We carried out simulations for $10^{-5}>E>10^{-6}$ for
model J and $10^{-5}>E>5\times10^{-6}$ for model N. The Rossby numbers are always smaller than $0.1$.
For the profile J, in cases of low Ekman numbers ($E\le 2\times 10^{-6}$), we imposed longitudinal symmetry by calculating only 
the harmonics of a chosen order $m_s$. 
The required resolution for $E=10^{-6}$ is $500$ points on the radial
grid and $l=580$ spherical harmonics degrees.

\subsection{Axisymmetric flow}
\begin{figure}
\centering
  \subfigure[Model J]{\label{fig:axi_J}
  \includegraphics[clip=true,width=0.75\textwidth]{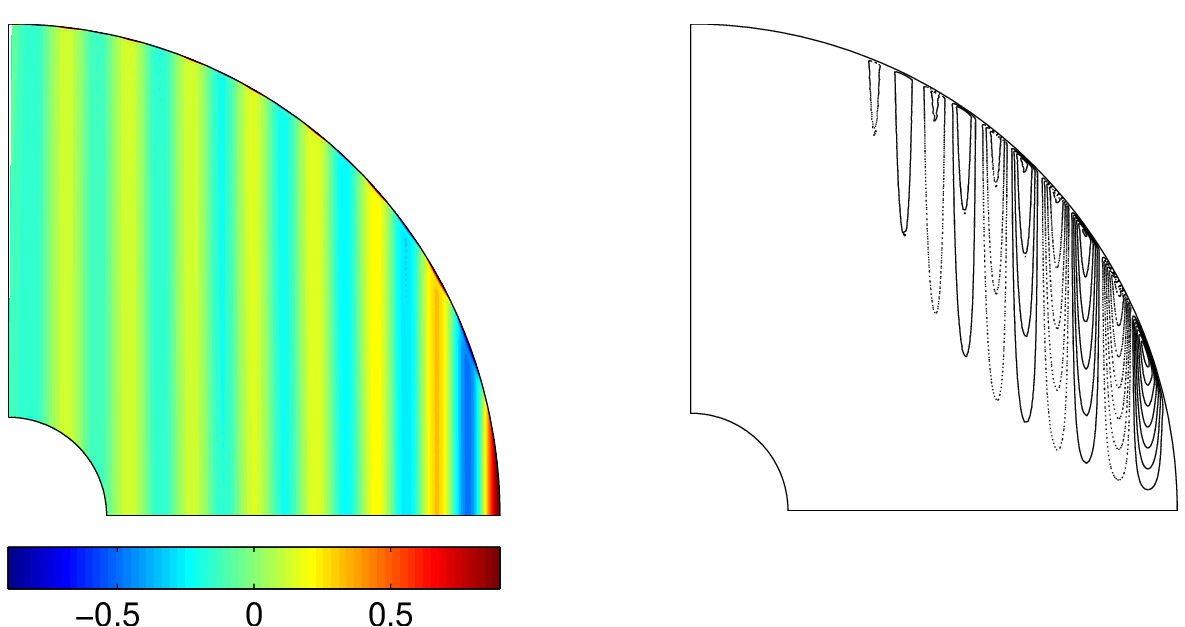}}
   \subfigure[Model N]{\label{fig:axi_N}
   \includegraphics[clip=true,width=0.75\textwidth]{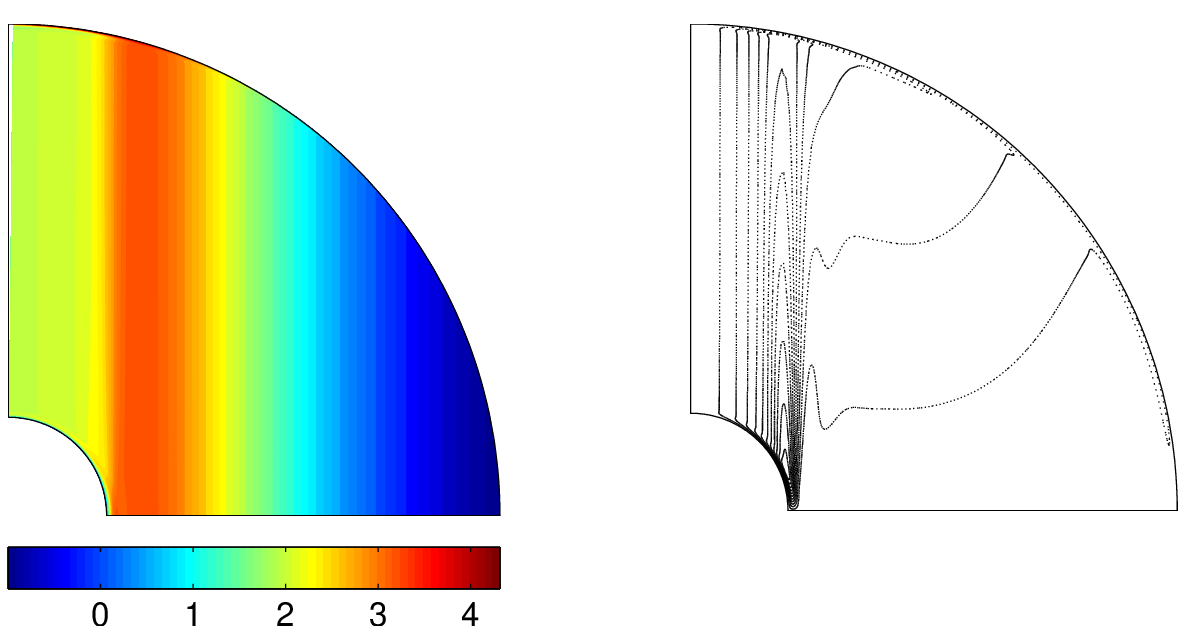}}      
\caption{Angular velocity $u_{\phi}/(r \sin \theta)$ (left) and streamlines of the meridional circulation 
(isocontours of $\psi=r \sin \theta \frac{\partial u_p}{\partial \theta}$ with $u_p$ the velocity poloidal scalar) (right) 
of the axisymmetric flow in the northern meridional plane. 
For the meridional circulation, anti-clockwise (clockwise) flows are shown in solid (dotted) lines. 
The parameter for the simulations are $E=5\times10^{-6}$ and $Ro=0.015$ for model J (a) 
and $E=10^{-5}$ and $Ro=0.02$ for model N (b).}
\label{fig:Uaxi_jupiter}
\end{figure}

When the imposed boundary forcing is small enough, \ie\ when the Rossby number $Ro$ is 
less than a critical value $Ro_c$, the flow is axisymmetric and predominantly azimuthal (figure~\ref{fig:Uaxi_jupiter}). 
The zonal jets imposed at the outer boundary extend 
into the volume along lines parallel to the axis of rotation. 

\setcounter{subfigure}{0}

\begin{figure}
\centering
  \subfigure[Model J]{\label{fig:profilJ_axi_eq}
   \includegraphics[clip=true,width=0.45\textwidth]{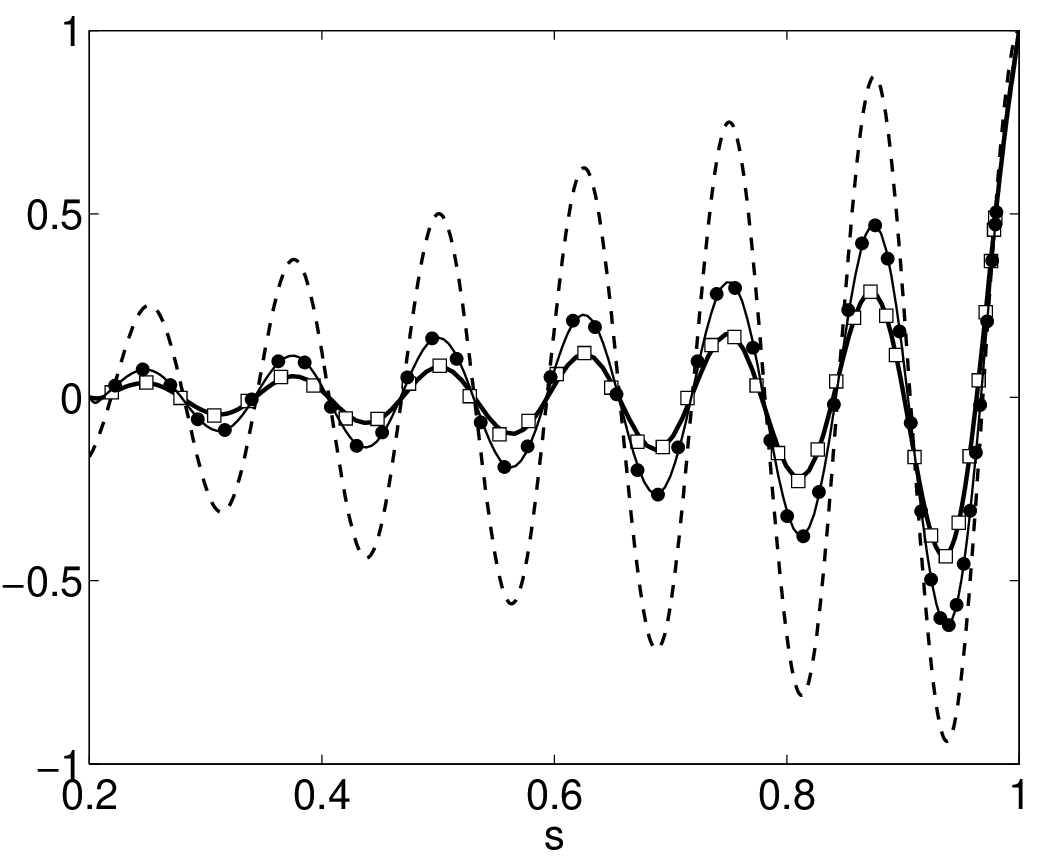}}
\hspace{0.5cm}
\subfigure[Model N]{\label{fig:profilN_axi_eq}
   \includegraphics[clip=true,width=0.45\textwidth]{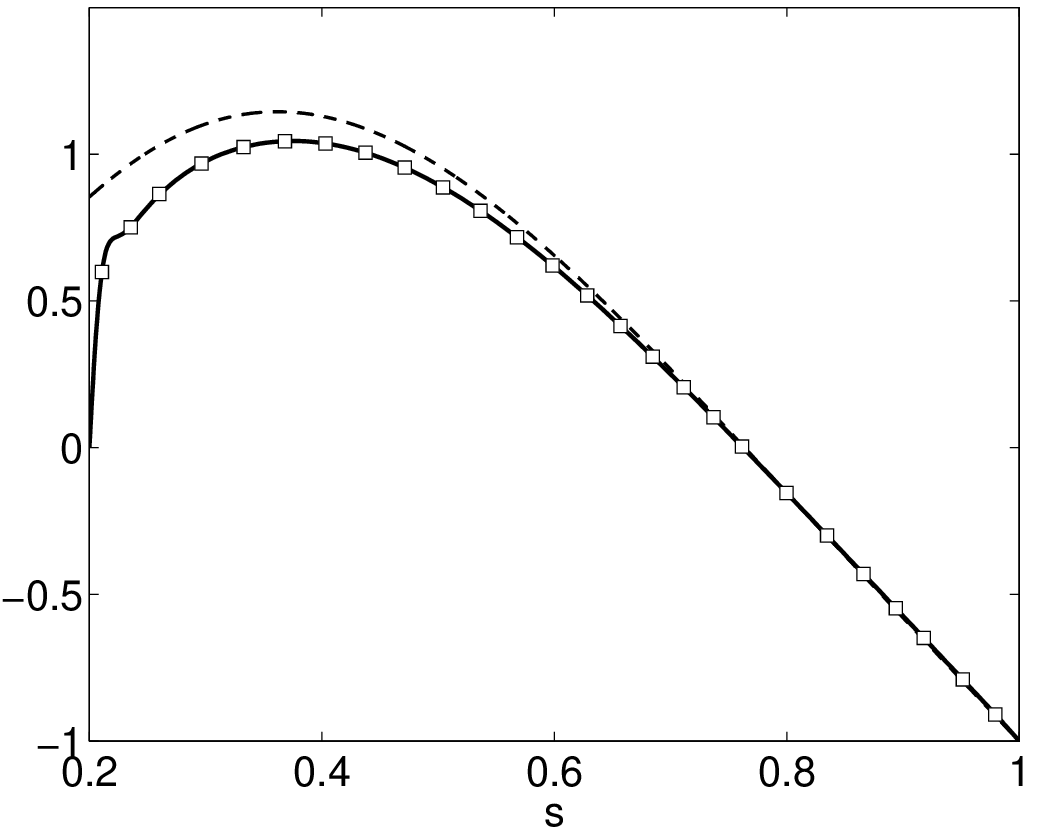}}
\caption{Zonal velocity in the equatorial plane for subcritical numerical simulations (solid lines) compared to the 
zonal velocity at radius $r=0.98 r_o$ (symbols) and imposed velocity at the top (dashed line) both projected in the equatorial plane for (a) model J ($E=5 \times 10^{-6}$ (bold solid line and open squares) and $E=10^{-6}$ (thin solid line and black circles)) and (b) model N ($E=10^{-5}$ (bold solid line and open squares)).}
\label{fig:Profil_axi_eq}
\end{figure}

The use of no-slip boundary conditions yields a differential rotation between the boundary and the bulk of the fluid. 
This differential rotation is accommodated across viscous Ekman boundary layers,
which scale as $(E/\cos \theta)^{1/2}$, where $\theta$ is the colatitude. 
By Ekman pumping, viscous forces within the Ekman layers drive axial motions of order $E^{1/2}$ within the bulk of the fluid (figure~\ref{fig:Uaxi_jupiter}). 
These meridional circulations advect angular momentum from the boundary layer into the bulk of the fluid
and cause the jets to propagate faster than by pure viscous diffusion. 
At low latitudes, the Ekman layer is thicker so the Ekman pumping is stronger,
yielding to a more efficient driving of the zonal motions in the bulk by the outer boundary layer.
For model J (figure~\ref{fig:profilJ_axi_eq}), the zonal velocity in the bulk relative to that
imposed at the outer boundary is noticeably weaker for the inner jets than for the outer jets. 
When $E$ decreases this effect is less marked, and in the $E\to 0$ limit we expect the 
basic zonal velocity to be perfectly geostrophic in the whole volume.
The comparison between the zonal velocity just below the Ekman layer and in the equatorial 
plane (figure~\ref{fig:Profil_axi_eq}) shows that the zonal velocity is geostrophic
in the bulk of the fluid (outside of the boundary layers). 
For model N (figure~\ref{fig:profilN_axi_eq}), the zonal jets are wider, so the zonal flow already 
displays a strong geostrophic structure at $E=10^{-5}$. 
Note that the azimuthal velocity has to match the no-slip boundary condition at the 
inner core, and so an internal Stewartson layer forms on the axial cylinder tangent to the inner core \citep{Ste66}.

\setcounter{subfigure}{0}

\subsection{Non-axisymmetric motions}
\subsubsection{Model J}
\paragraph{Rossby wave at the onset}

\begin{figure}
   \centering
   \includegraphics[clip=true,width=0.4\textwidth]{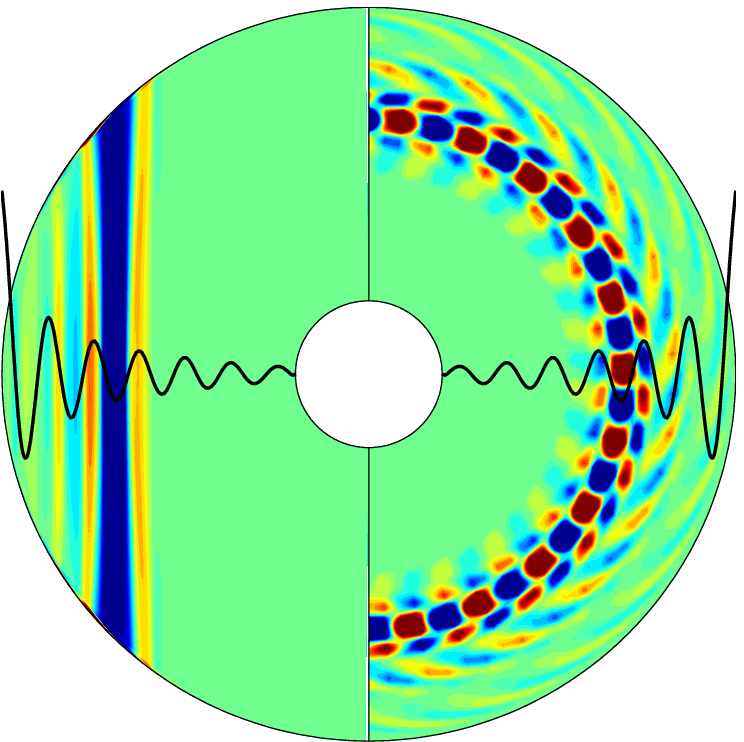}
    \caption{Non-zonal axial vorticity in the equatorial plane (right) and in a meridional slice 
   (left) for model J at $E=4\times 10^{-6}$ and $Ro=1.01 Ro_c$ 
   (blue: negative and red: positive). 
   The black curve represents the zonal velocity in the equatorial plane.}
\label{fig:W_Rec_J}
\end{figure}

When the boundary forcing (measured by $Ro$) becomes greater than a critical value $Ro_c$, the axisymmetric 
basic flow becomes unstable to a non-axisymmetric shear instability. 
The saturated instability takes the form of an azimuthal necklace of cyclonic and anticyclonic vortices
aligned with the axis of rotation, is nearly $z$-independent and drifts eastward (figure~\ref{fig:W_Rec_J}). 
Close to the threshold, the radial extension of the pattern is large and occupies almost 
half of the gap. The pattern drifts with the same speed over its whole
radial extension, even though the advection by the zonal flow velocity 
varies with $s$, implying that it is a single 
wave.

\citet{Wic02b} studied the linear stability of the imposed zonal flow~(\ref{eq:profileJ}) in 
a spherical shell modeling the insulating molecular hydrogen layer of Jupiter (aspect ratio 0.8). For $E=10^{-4}$ they found 
nearly bidimensional instabilities that they described as drifting columns aligned with the
rotation axis and similar to convective solutions. Although they do not identify these instabilities 
as waves, their characteristics are very similar to the ones obtained with our non-linear model.

The nearly $z$-invariant structure and the prograde drift are two characteristics of Rossby waves
propagating in a spherical container.
The dispersion relation for the Rossby wave given by a local linear analysis is \citep[\eg][]{Fin08}
\begin{eqnarray}
 \omega_{rw}(s)= -2 \Omega \beta \frac{m/s}{k_s^2+(m/s)^2} ,
\label{eq:w_RW}
\end{eqnarray}
where $\beta=h^{-1}(dh/ds)=-s/(r_o^2-s^2)$ is related to the slope of the upper boundary of the spherical container 
of height $h$. $k_s$ and $m/s$ are 
the radial and azimuthal wavenumbers respectively.
The theoretical Rossby wave frequency $\omega_{rw}$ can be calculated at a given radius 
assuming $k_s\approx m/s$
and using the wavenumber $m$ obtained from the numerical simulation.
For different $E$, the frequency $\omega$ of the propagating wave observed in our 
numerical simulations always falls in the range $\omega_{rw}(s_1)<\omega<\omega_{rw}(s_2)$ where $s_1=0.56$ 
($s_2=0.87$) is the smallest (resp. largest) radius where a
significant vorticity associated with the presence of the wave can be seen in the numerical calculations.
This strongly indicates that the shear instability occurs as a Rossby wave. 

The velocity of the zonal flow $U$ enters the dispersion relation of the Rossby wave through a Doppler shift
\begin{eqnarray}
 \omega(s)=\omega_{rw}(s)+U(s)\frac{m}{s} .
\end{eqnarray}
As reported earlier, $\omega(s)$ is constant in our numerical calculations so $\omega_{rw}(s)$ must adapt 
in the $s$-direction for the wave to be coherent. In a prograde jet $U>0$, $\omega_{rw}$ must decrease, 
which requires a local increase in $k_s$ in equation~(\ref{eq:w_RW}) and so a local decrease in the radial lengthscale, which can 
be observed in figure~\ref{fig:W_Rec_J}.
For small enough Ekman number (in practice $E < 5 \times 10^{-6}$), the critical wavenumber $m_c$ of the Rossby mode is
independent of $E$. The radial lengthscale is determined by the width of the jet and the vortices are roughly circular 
in the equatorial plane (figure \ref{fig:W_Rec_J}) suggesting that $m_c$ is controlled by the width of the jets. 

In a local approximation that neglects the curvature terms, a criterion of instability of barotropic shear flows 
has been derived by \citet{Ing82} for an anelastic model in a full rotating sphere and by \citet{Kuo49} for thin stably stratified ``weather'' layers.
Using an inviscid Boussinesq model and for barotropic instability of a zonal flow $U$ in a sphere, 
this necessary condition implies a change of sign of a quantity $\Delta$ at some radius:
\begin{eqnarray}
 \Delta = 2 \beta - Ro \frac{d \zeta}{ds},
\label{eq:crit_SC}
\end{eqnarray}
where $\zeta$ is the vorticity of the zonal flow,
\begin{eqnarray}
 \zeta = \frac{dU}{ds}+\frac{U}{s}.
\end{eqnarray}
Note that the curvature terms have been taken into account here.
In a sphere, $\beta$ is negative.
Consequently, the zonal velocity profile is more prone to instability where the gradient of zonal vorticity is maximum and negative. 
Then for a profile $U$ of sinusoidal form, the first shear instability occurs at the maximum 
of the prograde jets, and thus, perhaps surprisingly, at a null value of the zonal velocity shear $dU/ds$. 
Note that our numerical simulations show instabilities with a large radial extent and with maximum amplitude located in a
retrograde zonal jet  (see figure~\ref{fig:W_Rec_J}), even though the local instability criterion predicts 
an onset in a prograde jet. This observation emphasizes that the local criterion does not predict the location of global saturated modes.

\begin{figure}
 \centering
  \includegraphics[clip=true,width=\textwidth]{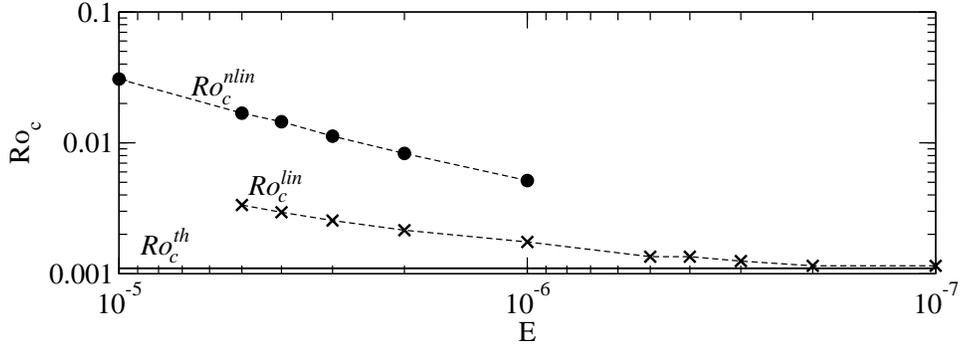}
  \caption{
  Critical Rossby number obtained from fully non-linear numerical simulations for model J ($Ro_c^{nlin}$, circles) 
  compared to the theoretical Rossby number obtained with the local instability criterion~(\ref{eq:crit_SC}) using the geostrophic profile~(\ref{eq:profileJ}) 
  $U(s,\theta=\pi/2)$ ($Ro_c^{th}$, black line). The critical Rossby number obtained from the linear numerical calculation is also shown 
  ($Ro_c^{lin}$, crosses).}
 \label{fig:Roc_E_J}
\end{figure}

The theoretical critical Rossby number obtained from applying the criterion~(\ref{eq:crit_SC}) to the profile~(\ref{eq:profileJ}) 
imposed at the top of model is $Ro_c^{th}=0.0011$.
The threshold of the first instability of the axisymmetric flow, denoted $Ro_c^{nlin}$, 
obtained with the numerical simulations are shown in figure~\ref{fig:Roc_E_J}. 
Despite the decrease of $Ro_c^{nlin}$ with the Ekman number, 
$Ro_c^{nlin}$ is still about four times larger than $Ro_c^{th}$ for $E=10^{-6}$ because the amplitude of the zonal
flow within the bulk is reduced by viscous boundary layers in the numerical simulations.
Due to computational limitations, it is not possible for us to carry out simulations at smaller $E$ with a fully 
non-linear code and prove the existence of an asymptotic regime for the inviscid instability threshold.
For this purpose we used a dedicated linear code described in \ref{app:codeXSHELL}.
The linear code calculates linear perturbation solutions to the momentum equation 
using the geostrophic profile $U$ as the basic flow in the bulk of the fluid.
The computational time is greatly reduced by the linear approach but is restricted to an analysis of the
instability threshold.
The growing solutions obtained with the linear code exhibit very similar features to the
Rossby waves in the non-linear simulations
(frequency, bidimensional structure, radial extent, location of the maximum amplitude in a retrograde jet).
In figure~\ref{fig:Roc_E_J} the threshold $Ro_c^{lin}$ obtained with the linear code 
approaches asymptotically the value given by the local theory.
For the same Ekman number, $Ro_c^{lin}$ is smaller
 than $Ro_c^{nlin}$ since the geostrophic zonal flow 
$U$ is used in the linear code, that is the jets in the bulk have greater amplitude than in the non-linear 
code. 
From our linear computations we conclude that the theoretical 
criterion~(\ref{eq:crit_SC}) is relevant to explain the onset of instability obtained numerically.
More details about the onset of the hydrodynamic instability can be found in \citet{Gue10_thesis}.

The characteristic time of the Rossby wave is $\tau_{rw}=1/\omega$. 
At the instability threshold, the numerical simulations give $\tau_{rw}\approx 18 \Omega^{-1}$ for $E< 5\times 10^{-6}$.
The timescale of the zonal jets is $\tau_{zj}=r_o/U_0=\Omega^{-1}/Ro$. 
For $Ro=0.01$, we have $\tau_{zj}>\tau_{rw}$: the Rossby wave propagation is faster than the advection 
of the fluid by the zonal flow. 
The turnover time of a fluid particle trapped in a Rossby wave is $\tau_{to}=l/V_s$ where $l$ is the typical radial displacement of the
particle and $V_s$ the typical cylindrical radial velocity of the particle. 
At $Ro=1.01Ro_c$, $V_s$ is typically $10^{-2} U_0$. 
In a rough approximation we use $l=\delta$, where $\delta$ is the width of the jets, $\delta \approx 0.1r_o$ for the
profile J. 
Then we obtain $\tau_{to} \approx 0.1 r_o/(10^{-2} U_0) \approx 10 Ro^{-1} \Omega^{-1} \approx 10^{3}\Omega^{-1}$: the turnover time of the particle is
much longer than the timescale of the wave. Consequently the particle oscillates rapidly as the wave propagates 
and is slowly advected by the zonal flow. In practice the radial displacement $l$ is typically smaller than $\delta$ and so
the turnover time is slightly overestimated here.

\paragraph{Supercritical regime}

\begin{figure}
\centering
  \subfigure[$Ro=1.06 Ro_c$]{\label{fig:U_E5e6R3600}
  \includegraphics[clip=true,width=0.75\textwidth]{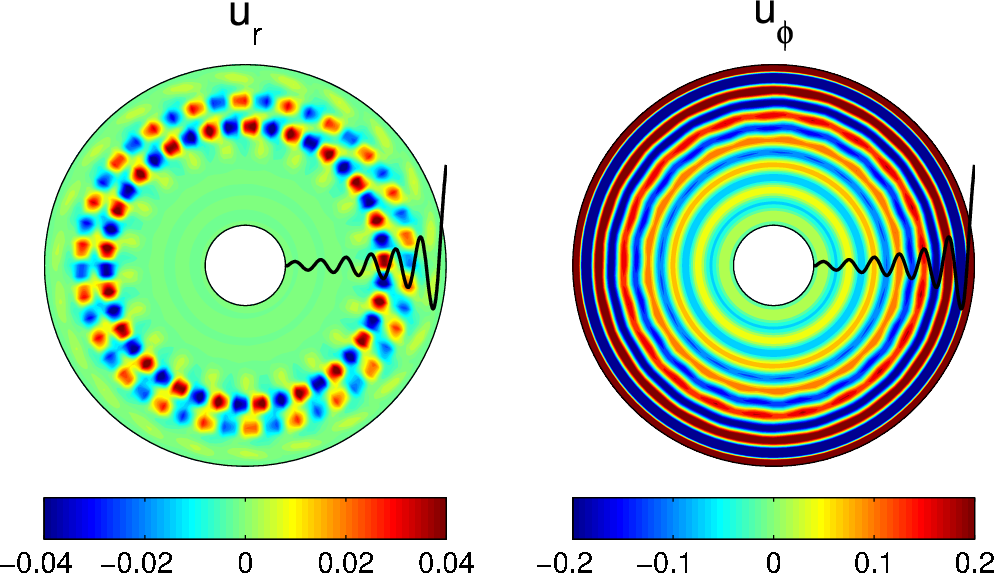}}   
\subfigure[$Ro=2.94 Ro_c$]{\label{fig:U_E5e6R10000}
 \includegraphics[clip=true,width=0.75\textwidth]{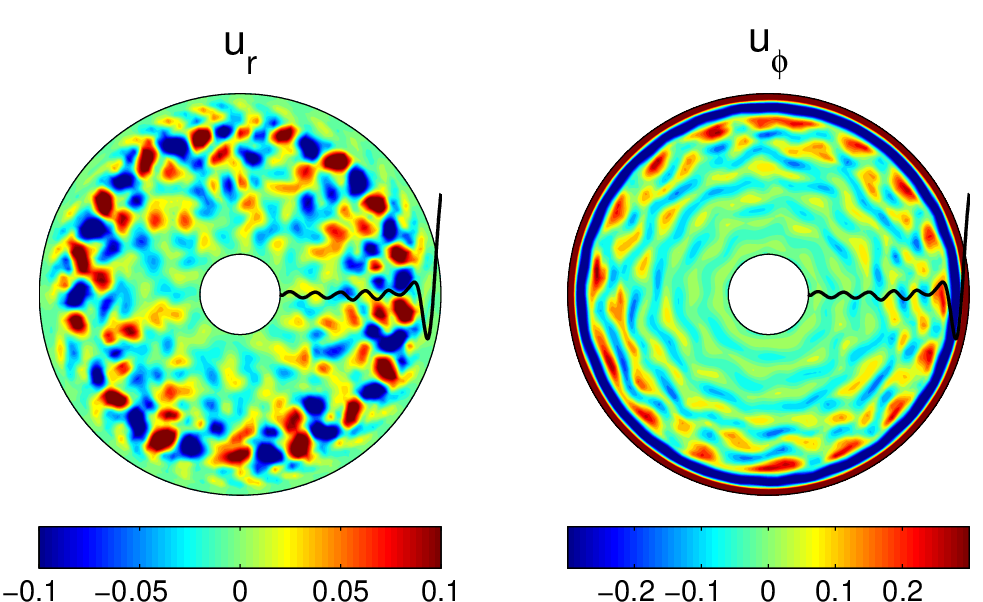}}  
\caption{Snapshots of the radial (left) and azimuthal (right) velocity components in the equatorial plane for $E=5\times 10^{-6}$ and $Ro>Ro_c$ for model J. The velocities are scaled by $U_0$. For $u_{\phi}$ the colorscale has been truncated ($u_{\phi}(r_o,\theta=\pi/2,\phi)=1$).
The black curve represents the zonal velocity in the equatorial plane.}
\label{fig:ProfJ_U_Rosurcrit}
\end{figure}

When the Rossby number is increased in the supercritical regime, other prograde 
jets will eventually become unstable.
A second Rossby wave appears in the weakly supercritical regime, at $Ro=1.06 Ro_c$ for $E=5\times 10^{-6}$,
with a maximum velocity located 
in the retrograde zonal jet at larger radius than the first wave maxima (\ie\ the wave appearing for $Ro=Ro_c$) (figure~\ref{fig:U_E5e6R3600}). 
To fill the larger circumference at larger radius the instability has a slightly larger wave number, $m=22$ instead of $21$, while the radial width of the jet is comparable. 
The second wave propagates faster, in agreement with the Rossby wave dispersion relation~(\ref{eq:w_RW}).
Barotropic instabilities tend to broaden and weaken narrow jets 
by redistributing potential vorticity \citep[see for instance ][]{Ped87}.
The smoothing of the jets saturates the amplitude of the Rossby waves.
For this slightly supercritical regime the zonal flow profile is only weakly modified.
Upon further increasing the forcing ($Ro=2.94 Ro_c$), 
several Rossby waves of different wavenumbers superpose and interact (figure~\ref{fig:U_E5e6R10000}).
The structure of the waves and the jets is still mainly bidimensional except in the viscous boundary layers.
The typical cylindrical radial velocity is $V_s \approx 0.1 U_0$ and the Rossby number is $0.05$ so the 
turnover time is about $20\Omega^{-1}$ assuming that the radial displacement $l=\delta$, about the same order of magnitude as the timescale of the zonal jets.

\setcounter{subfigure}{0}

\begin{figure}
\centering
  \subfigure[]{\label{fig:U_s_satur}
   \includegraphics[clip=true,width=0.43\textwidth]{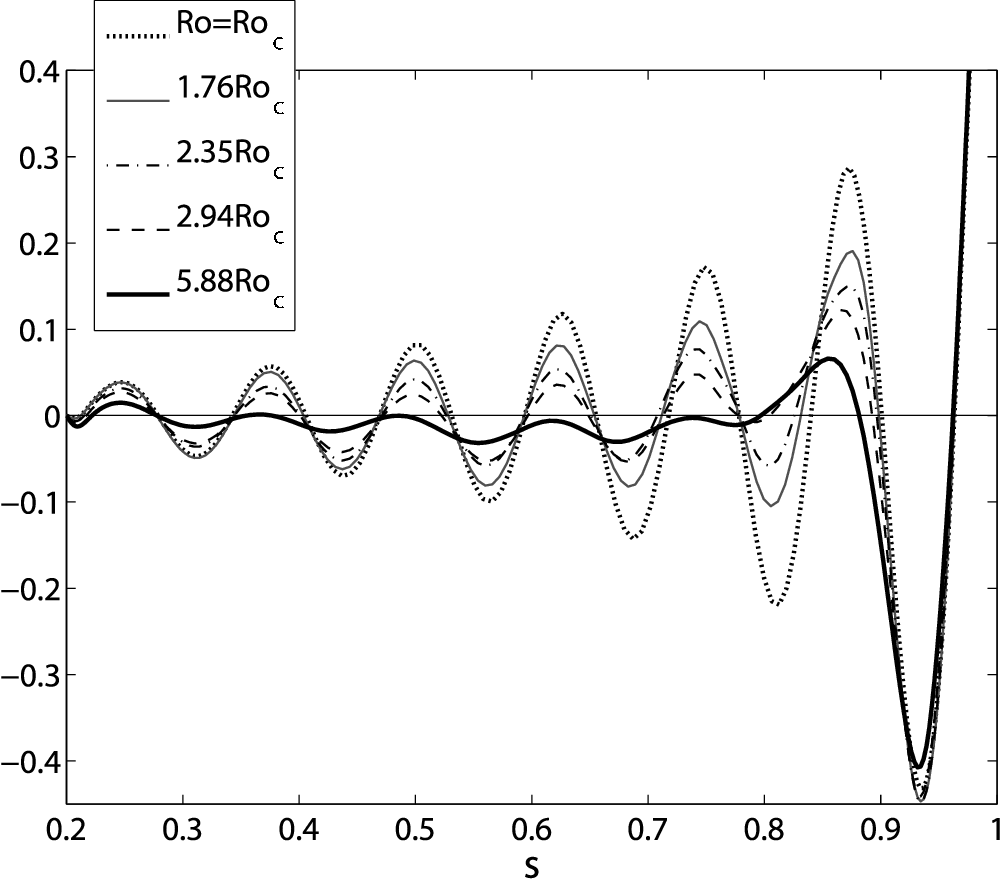}}
\subfigure[]{\label{fig:U_Roc_satur}
   \includegraphics[clip=true,width=0.47\textwidth]{fig9b.eps}}
\caption{(a) Time-averaged zonal velocity in the equatorial plane for $E=5\times 10^{-6}$ and different
forcings. (b) Amplitude of the non-axisymmetric radial velocity $V_s$ (squares), non-axisymmetric azimuthal 
velocity $V_{\phi}$ (circles), non-axisymmetric velocity $(V_s^2+V_{\phi}^2)^{1/2}$ (diamonds) and zonal velocity at the radius $s=0.75$ (triangles). 
All velocities were measured in the equatorial plane in the units of $U_0$.
The amplitude of the non-axisymmetric velocity corresponds to the maximum in a snapshot, whereas the zonal velocity amplitude has been averaged in time.}
\label{fig:U_satur}
\end{figure}

In figure~\ref{fig:U_s_satur} the time-averaged zonal flow in the equatorial plane 
is plotted for different $Ro$ up to $Ro=5.88 Ro_c$. 
As the forcing is increased, the Rossby waves gradually reduce the jet strength and broaden the jet width. 
For $Ro=2.94 Ro_c$, the retrograde jet at $s=0.81$ has been
mostly destroyed leading to the widening of the zonal jet width. We note that the zonal flow becomes mostly westward for the strongest forcings.
The amplitude of the zonal flow located at $s>0.9$ is hardly affected 
because the threshold to destabilise the outermost jets is high due to the large slope (related to $\beta$ in equation~(\ref{eq:crit_SC})).
For $Ro<2.35 Ro_c$, the amplitude of the non-axisymmetric velocity, relative to $U_0$, increases with the forcing (figure~\ref{fig:U_Roc_satur}).
After reaching a maximum, at $Ro=2.35 Ro_c$, the amplitude of the non-axisymmetric flow decreases relative to $U_0$.  
The ``efficiency'' of the forcing to drive the non-zonal velocity is reduced 
as the Rossby waves smooth the gradient of vorticity and so affect their excitation mechanism.

The back reaction on the forcing velocity in the upper molecular hydrogen layer is not taken into account in our 
model although it might significantly affect the zonal profile in the upper layer in the case of strong forcing.

\setcounter{subfigure}{0}

\subsubsection{Model N}
\begin{figure}
\centering
  \includegraphics[clip=true,width=.4\textwidth]{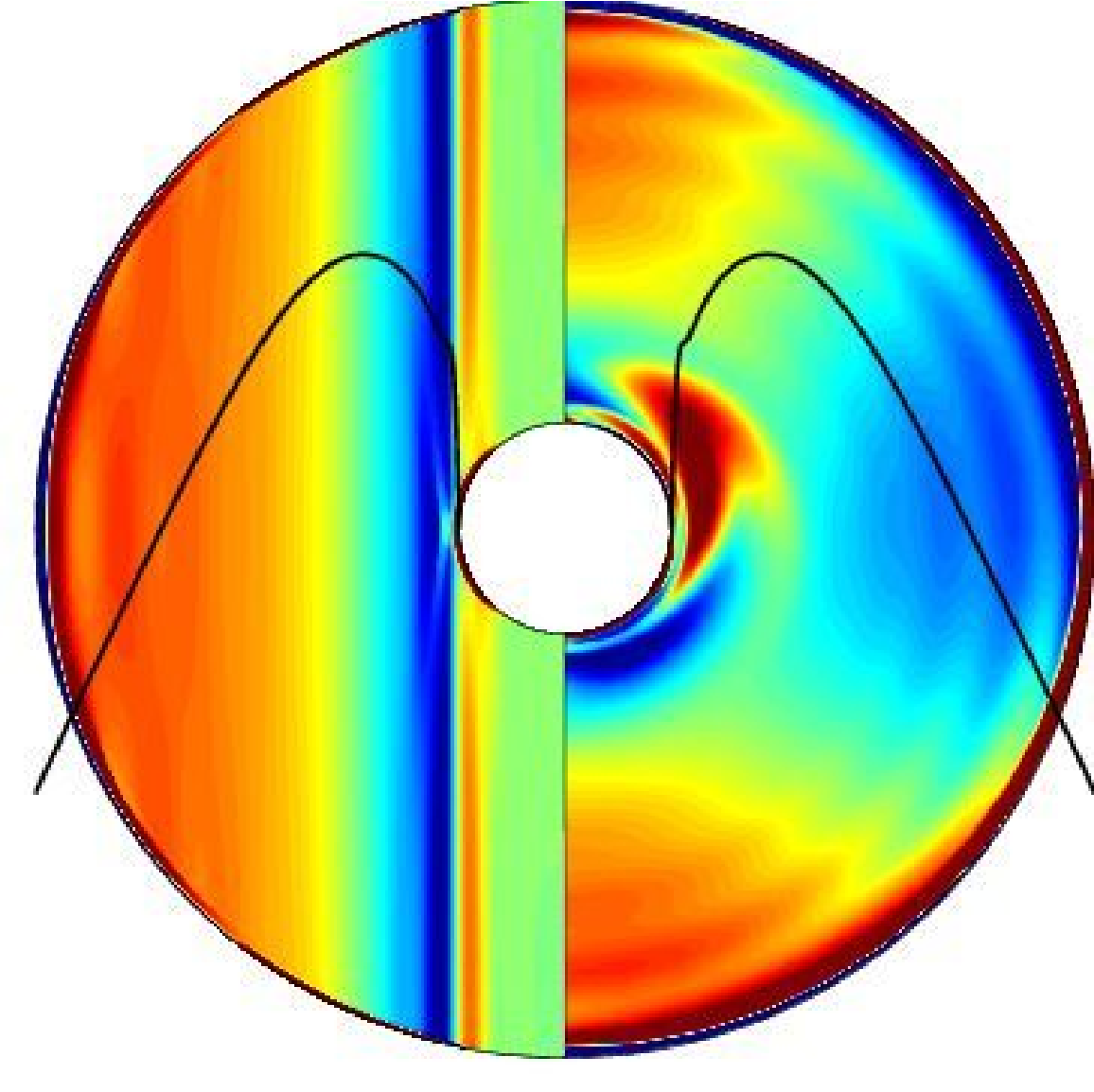}    
   \caption{Non-zonal axial vorticity in the equatorial plane (right) and in a meridional slice (left) for model N at $E=5\times 10^{-6}$ and $Ro=1.01Ro_c$ 
   (blue: negative and red: positive). The black curve represents the zonal velocity in the equatorial plane.}
   \label{fig:W_Rec_N}
\end{figure}

The shear instability takes the form of an $m=2$ oscillation in the azimuthal direction (figure~\ref{fig:W_Rec_N}).
It is a single wave propagating eastward with the same frequency over the shell, and
is nearly $z$-invariant. The maxima of the non-zonal vorticity are located on each side of the prograde 
jet. The characteristics of this wave are similar to the Rossby wave obtained with model J. 
The frequency of this wave is in agreement with the frequency of a theoretical Rossby wave of wavenumber 
$m=2$ propagating at a radius $s=0.53$ (assuming that $k_s\approx m/s$ in the dispersion relation~(\ref{eq:w_RW})).
For $E=10^{-5}$ and $E=5\times 10^{-6}$, the critical Rossby numbers obtained with the non-linear numerical simulations are respectively
$Ro_c^{nlin}=0.0335$ and $Ro_c^{nlin}=0.0325$.  
Using the instability criterion~(\ref{eq:crit_SC}) with the profile imposed at the surface we obtain a critical Rossby number of $0.026$ in 
good agreement with the non-linear numerical results when the Ekman number decreases. 

\section{Magnetic field generation}
\label{sec:dynamo_jupiter}
The non-axisymmetric motions are of prime importance for the dynamo mechanism because a purely 
toroidal flow cannot generate a self-sustained magnetic field. We note that some axisymmetric poloidal flow is
present when $Ro<Ro_c$ as a weak meridional circulation is created by the Ekman pumping. 
However these axisymmetric motions are weak at small Ekman numbers so we do not expect
to find dynamos when the zonal flow is stable, that is when $Ro<Ro_c$, in the asymptotic inviscid regime.
Indeed we did not find dynamos when $Ro<Ro_c$ (up to $Pm=10$). 
The non-axisymmetry associated with the hydrodynamic shear instability is a crucial element for the dynamo process: the stable zonal flow cannot sustain a magnetic field by itself. This is in agreement with the results obtained by \citet{Gue10} with dynamos generated by spherical Couette flows (differential rotation between two concentric spheres).

\subsection{Characteristics of the magnetic field for model J}
We have performed dynamo simulations for $Ro=1.17-1.76Ro_c$ and $E=5\times 10^{-6}$. 
We find that the dynamo threshold occurs at a rather high value of the magnetic Prandtl number, \mbox{$Pm_c\approx5$}. 
The critical
magnetic Reynolds number (defined via the maximum forcing velocity) required for dynamo action is $Rm_c\approx20,000$ (see section~\ref{sec:Summary} for an estimate of the critical magnetic Reynolds number defined via the local velocity). 
For a given forcing, we have performed calculations just above the critical magnetic Prandtl number, $Pm_c$, and up to $2Pm_c$.

\begin{figure}
\centering
  \subfigure[Axisymmetric magnetic field]{\label{fig:Baxi_ProfJ}
  \includegraphics[clip=true,width=0.4\textwidth]{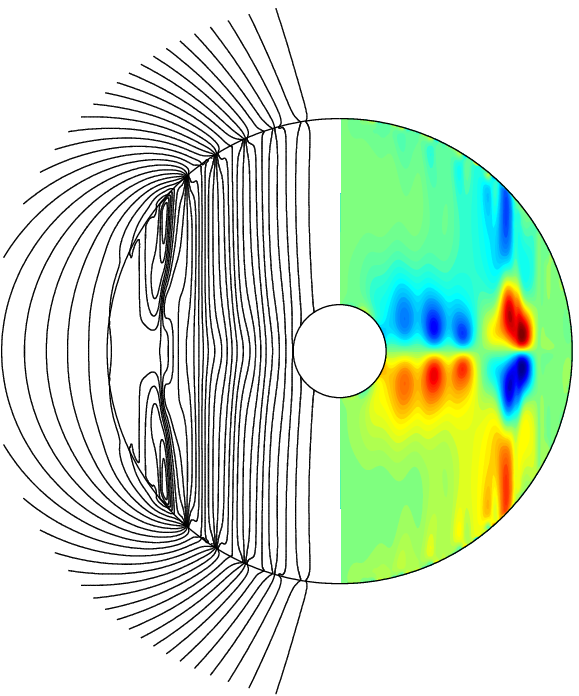}}
  \subfigure[Radial magnetic field at $r_s$]{\label{fig:Bmap_rs_J}
  \includegraphics[clip=true,width=0.7\textwidth]{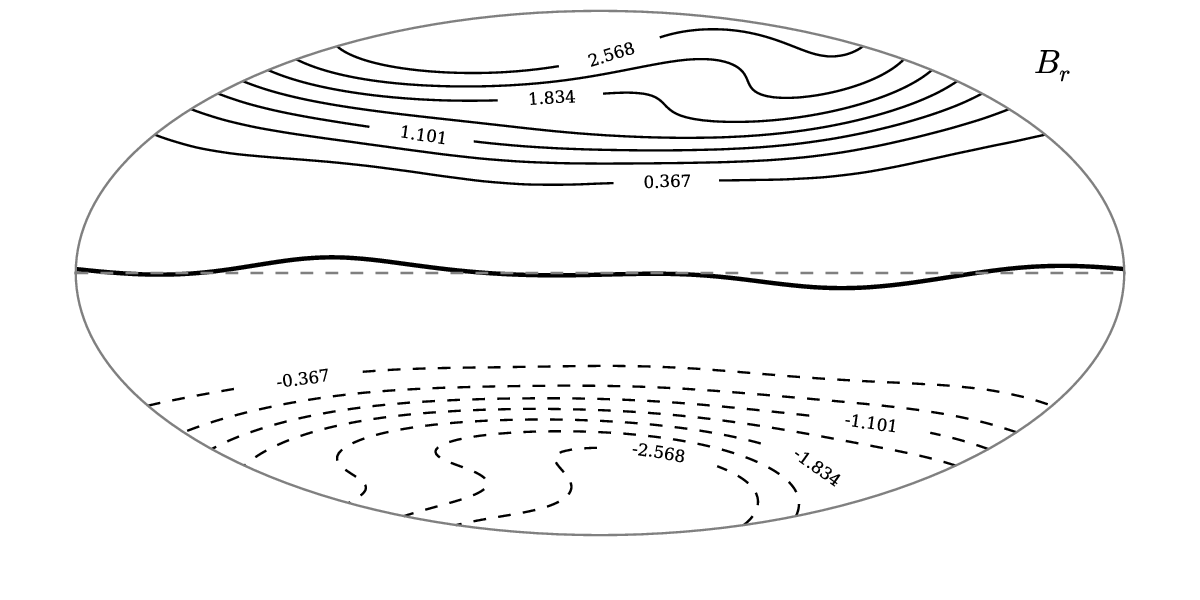}}
 \caption{Magnetic field for model J. 
(a) Snapshot of the axisymmetric magnetic field in a meridional plane: magnetic poloidal field lines (left) and azimuthal magnetic field (right) 
(blue: negative and red: positive).
(b) Map of the radial magnetic field at the surface of the planet $r_s=1.25r_o$ in unit of $10^{-3} \sqrt{\rho \mu_0}U_0$
(solid line: positive and dotted line: negative).
The poloidal magnetic field at $r=r_s$ is calculated assuming the region between $r_o$ and $r_s$ is electrically insulating.
The parameters of this simulation are $E=5\times10^{-6}$, $Ro=1.17Ro_c$ and $Pm=5 \approx Pm_c$.}
\label{fig:B_ProfJ}
\end{figure}
\setcounter{subfigure}{0}

\begin{figure}
\centering
  \subfigure[Kinetic and magnetic energies in the fluid]{\label{fig:spec_fluid_J}
   \includegraphics[clip=true,width=0.7\textwidth]{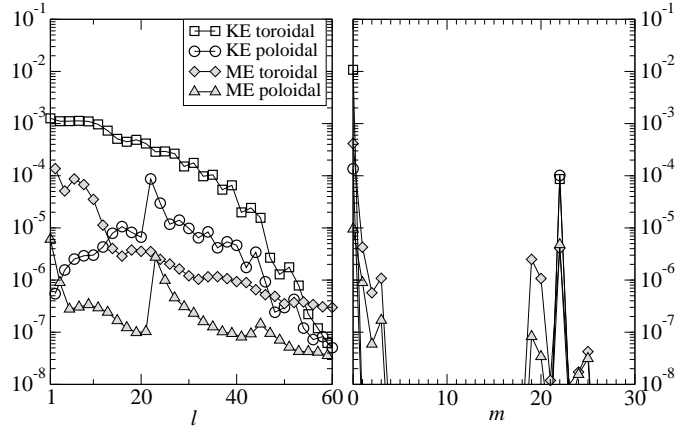}}
  \subfigure[$A_l$ and $A_m$ at $r_s$]{\label{fig:spec_rs_J}
   \includegraphics[clip=true,width=0.7\textwidth]{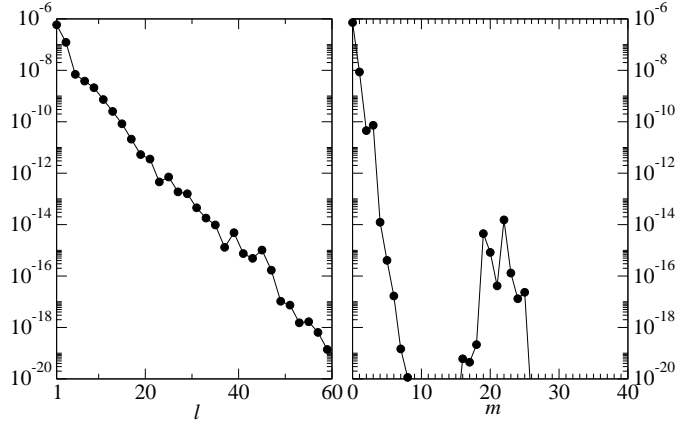}}
\caption{Magnetic energy spectra for model J. 
(a): Kinetic (KE) and magnetic (ME) energy per unit volume for each spherical harmonics degree $l$ (left) and mode $m$ (right) in the fluid conducting region given in unit of $\rho U_0^2$.
(b): Squared amplitudes of the magnetic field, $A_l$ (left) and $A_m$ (right) as defined in figure~\ref{fig:mag_spec}, at $r_s=1.25r_o$ given in unit of $\rho \mu_0 U_0^2$. 
Only the degrees of significant amplitude have been plotted, that is, $l$ even for the kinetic poloidal and
magnetic toroidal energies and $l$ odd for the kinetic toroidal and magnetic poloidal energies.
These data are taken at a particular instant and have not been time-averaged.
Same parameters than figure~\ref{fig:B_ProfJ}.}
\label{fig:specB_ProfJ}
\end{figure}
\setcounter{subfigure}{0}

The main features of the self-sustained magnetic field can be observed in figures~\ref{fig:B_ProfJ} and \ref{fig:specB_ProfJ}.
The magnetic field displays a dipolar symmetry, \ie\ antisymmetry with respect to
the equatorial plane,
\begin{eqnarray}
(B_r,B_{\theta},B_{\phi})(r,\pi-\theta,\phi)=(-B_r,B_{\theta},-B_{\phi})(r,\theta,\phi) .
\label{eq:sym_dip}
\end{eqnarray} 
The magnetic field is predominantly toroidal and axisymmetric (corresponding to the mode $m=0$ in
figure~\ref{fig:spec_fluid_J}). The toroidal magnetic field does not emerge from the conducting region as the 
outer region is electrically insulating.
The strongest poloidal component is the axial dipole within the conducting region and outside of the outer sphere
(corresponding to the harmonic $(l,m)=(1,0)$ in figure~\ref{fig:specB_ProfJ}). 
Within the bulk of the flow, the axisymmetric poloidal magnetic field lines are mostly significantly bent where
the Rossby wave causes a strong magnetic induction (figure~\ref{fig:Baxi_ProfJ}). 
A magnetic field at the scale of the Rossby wave is produced in this region as can be observed on the 
spectra of magnetic energy (figure~\ref{fig:spec_fluid_J}) with significant peaks at $m=22$ in 
the poloidal and toroidal magnetic energies and at $l=23$ in the poloidal magnetic energy ($l-m$ is odd to 
preserve the dipolar symmetry).
Close to the outer boundary, the axisymmetric poloidal magnetic field lines converge and diverge locally 
(figure~\ref{fig:Baxi_ProfJ}). This is due to 
the induction of axisymmetric magnetic field by the secondary meridional circulation produced by 
Ekman pumping. This effect is very localized and generates a magnetic field of small latitudinal scale 
that decreases rapidly with radius.

The spectrum and map of the radial magnetic field at the surface of our modeled planet (at radius $r_s=1.25r_o$) (Figs.~\ref{fig:Bmap_rs_J} and \ref{fig:spec_rs_J}) 
show that the magnetic field is strongly dominated 
by the axial dipole. 
The magnetic field generated at the scale of the Rossby wave ($m=22$) is still visible in
the spectrum of the magnetic field 
but its amplitude is weak at this radius: about four orders of magnitude smaller than the amplitude 
of the axisymmetric mode (note that the spectrum in figure \ref{fig:spec_rs_J} represents the squared amplitude of 
the field). 

In all the simulations performed, no inversion of polarity of the axial dipole has been observed. 
The tilt of the dipole is rather weak, at most $2^{\circ}$ from the rotation axis.
We found a secular variation of the dipole axis of about $1^{\circ}$ every $1000$ rotation periods
or alternatively $0.001$ global magnetic diffusion time. 

Just above the dynamo threshold ($Pm_c<Pm\leqslant 2Pm_c$), the magnetic field is weak: the magnetic energy contained 
within the fluid conducting region is only about $5$\% 
of the kinetic energy. The magnetic field does not strongly act back on the flow, except to produce its own saturation. 
A comparison between the zonal flow in the non-magnetic case and in the presence of the dynamo magnetic field does not reveal significant differences. 
The magnetic field lines of the poloidal field are almost aligned with the rotation
axis and the flow structure (see figure~\ref{fig:Baxi_ProfJ}) so the flow disruption due to Lorentz forces is weak.

\subsection{Characteristics of the magnetic field for model N}
\label{sec:B_profN}
We performed simulations at $Ro=1.05-1.5 Ro_c$ and $E=10^{-5}$. We find the dynamo 
threshold at $Pm_c\approx1$, that is, the critical magnetic Reynolds number is $Rm_c\approx4000$. 

\begin{figure}
\centering
 \subfigure[Axisymmetric magnetic field]{\label{fig:Baxi_ProfN}
 \includegraphics[clip=true,width=0.4\textwidth]{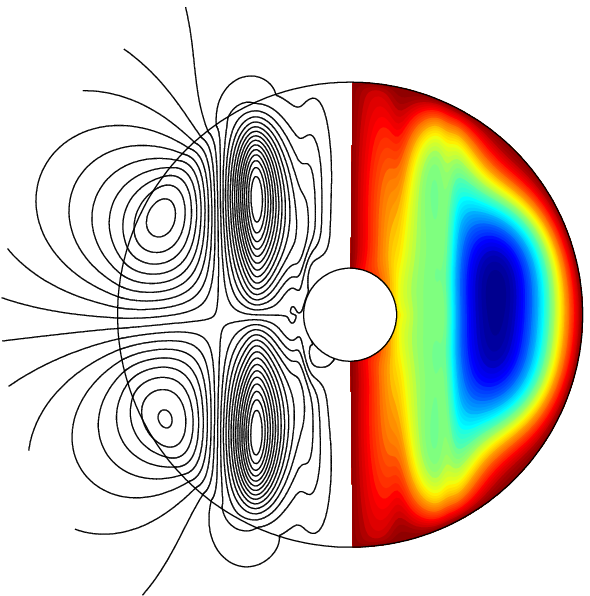}} 
\subfigure[Radial magnetic field at $r_s$]{\label{fig:Bmap_rs_N}
   \includegraphics[clip=true,width=0.7\textwidth]{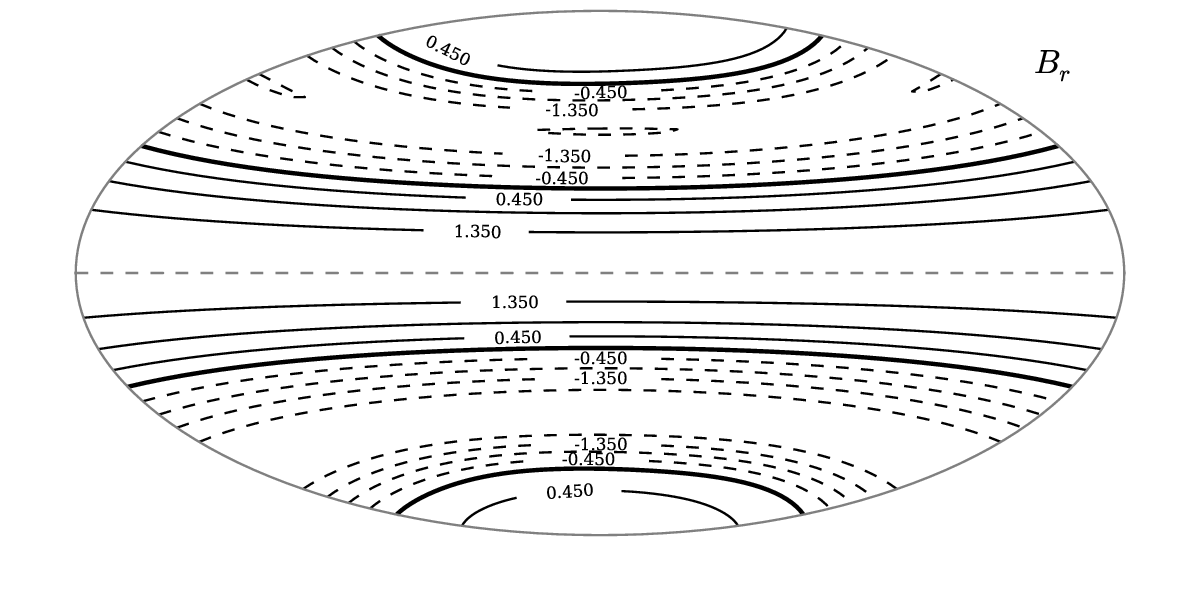}}
\caption{Magnetic field for model N (same as figure~\ref{fig:B_ProfJ}).
For the axisymmetric azimuthal field, blue corresponds to negative values and red to zero values.
The radial magnetic field is plotted at the surface of the planet $r_s=1.18r_o$ in unit of $10^{-5}\sqrt{\rho \mu_0}U_0$.
The parameters of this simulation are $E=10^{-5}$, $Ro=1.20Ro_c$ and $Pm=2 \approx2 Pm_c$.}
\label{fig:B_ProfN}
\end{figure}
\setcounter{subfigure}{0}

\begin{figure}
\centering
  \subfigure[Kinetic and magnetic energies in the fluid]{\label{fig:spec_fluid_N}
   \includegraphics[clip=true,width=0.7\textwidth]{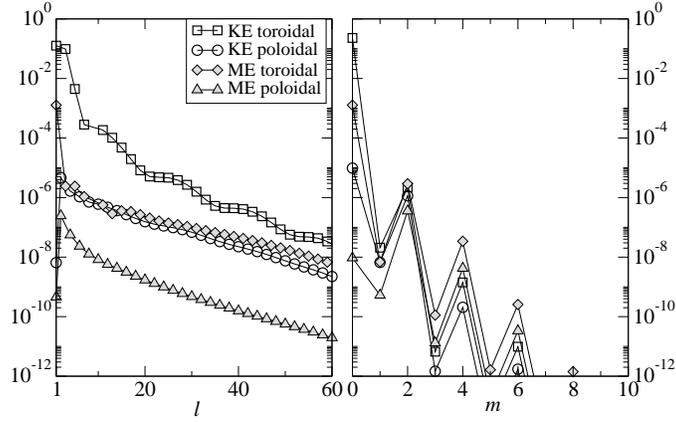}}
  \subfigure[$A_l$ and $A_m$ at $r_s$]{\label{fig:spec_rs_N}
   \includegraphics[clip=true,width=0.7\textwidth]{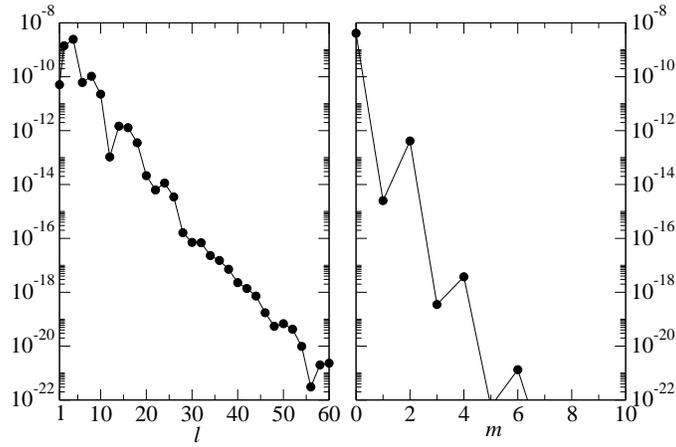}}
\caption{Magnetic energy spectra for model N (same as figure~\ref{fig:specB_ProfJ}). 
Only the degrees of significant amplitude have been plotted, that is, $l$ even for the kinetic poloidal and
magnetic poloidal energies (plus $l=1$) and $l$ odd for the kinetic toroidal and magnetic toroidal energies.
Same parameters than figure~\ref{fig:B_ProfN}.}
\label{fig:specB_ProfN}
\end{figure}
\setcounter{subfigure}{0}

The main features of the self-sustained magnetic field can be observed in figures~\ref{fig:B_ProfN} and \ref{fig:specB_ProfN}.
The self-sustained magnetic field displays an equatorial symmetry, \ie\
\begin{eqnarray}
(B_r,B_{\theta},B_{\phi})(r,\pi-\theta,\phi)=(B_r,-B_{\theta},B_{\phi})(r,\theta,\phi) .
\label{eq:sym_quad}
\end{eqnarray}
Within the fluid conducting region, the axisymmetric toroidal field is the strongest component whereas the poloidal field 
is dominated by the $m=2$ mode, not the axisymmetric $m=0$ mode. The $m=2$ mode
corresponds to the magnetic field generated at the scale of the Rossby wave (figure~\ref{fig:spec_fluid_N}).
The axisymmetric poloidal field is multipolar, mainly composed by the $(l,m)=(2,0)$ (axial quadrupole) and 
$(l,m)=(4,0)$ modes. At the surface of the planet (figure~\ref{fig:spec_rs_N}), the magnetic field appears to
be mainly axisymmetric (with the $l=2$ and $l=4$ harmonics degrees dominant). The amplitude of the $m=2$ structure is weak, 
about two orders of magnitude smaller than the $m=0$ mode but still visible at high latitudes 
on the map of the radial field at the surface (figure~\ref{fig:Bmap_rs_N}).

Due to the equatorial symmetry of the field, the magnetic field lines in the equatorial plane are roughly perpendicular
to the cylindrical structure of the flow whereas they are nearly aligned at higher latitudes (figure~\ref{fig:Baxi_ProfN}). 
As a result magnetic braking acting on the flow is stronger in the equatorial region than 
at high latitude regions. In the simulation performed here ($Pm_c<Pm\leqslant 2Pm_c$), the magnetic energy is weak 
compared to the kinetic energy (about $5$\%) so the feedback of the magnetic field on the flow remains weak. 
For a stronger magnetic field (at larger magnetic Reynolds numbers), we expect that the flow disruption would become 
important. As a result the equatorially symmetric solution may become unstable and the magnetic field may switch to 
an axial dipolar symmetry. This is the result obtained by \citet{Aub04} in convectively-driven dynamos: they found
 equatorial dipolar magnetic fields for Rayleigh numbers close to the convection onset; these solutions become 
unstable as the convective forcing is increased and an axial dipolar configuration is preferred.

In summary, the flows driven by the profiles J and N produce very different poloidal magnetic fields:
mainly a strongly axisymmetric dipole for the profile J and a weak multipolar axisymmetric field dominated by
the magnetic field induced at the scale of the Rossby waves for the profile N.
In both cases the magnetic field within the conducting region is mainly an axisymmetric toroidal field.
The different magnetic field morphology is quite surprising given that the flows are quite similar:
strong zonal flows and propagating Rossby waves.
In the next section we review the dynamo mechanism that has been proposed to operate for similar flows and 
suggest the key difference between profiles J and N that determines the topology of their self-sustained magnetic fields.

\subsection{Dynamo mechanism}
Using a quasi-geostrophic flow and a kinematic approach (no Lorentz force in the momentum equation),
\citet[][hereafter SC06]{Sch06} obtain numerical dynamos generated by an unstable axisymmetric shear layer (Stewartson layer): 
for a strong enough forcing, the Stewartson 
layer is unstable to non-axisymmetric shear instabilities, which appear in the form of Rossby waves (of wavenumber about $10$
for the Ekman numbers and Rossby numbers they investigated). 
The self-sustained magnetic field has a strong axisymmetric toroidal component and a mostly 
axisymmetric poloidal component. 
SC06 show that the time dependence of the flow is a key ingredient for the dynamo effect: time-stepping 
the magnetic induction equation using a steady flow taken either from a snapshot or a time-average leads 
to the decay of the magnetic field. 
They characterize the dynamo process as an $\alpha \omega$ mechanism.
In mean field theory, the $\alpha$ effect parameterizes the generation of an axisymmetric poloidal magnetic field 
from the correlation of small scale magnetic field and velocity. 
The $\alpha$ effect usually requires that the flow possess some helicity, the correlation between fluid velocity and vorticity, 
$H=\vel \cdot \boldsymbol{\omega}$ \citep[\eg][]{Mof78}. Flows displaying a columnar structure aligned with the axis of rotation, 
such as Rossby waves or convection columns \citep{Ols99}, typically possess strong mean helicity.
As these columns are essentially bidimensional vortical structures, the helicity is mainly produced by the term $u_z \omega_z$. 
In nearly $z$-invariant flow, the axial ($z$) velocity is mostly due to two terms: the slope effect and the Ekman pumping. 
The slope effect comes from the combination of mass conservation and impenetrable boundaries: a (cylindrical) radial velocity $u_s$ 
creates an axial velocity $u_z\sim z \beta u_s$ with $\beta=h^{-1} (dh/ds)$. In the limit of rapid rotation in a spherical container, 
this contribution is much larger (of order $1$) than the Ekman pumping  (\mbox{$u_z \sim E^{1/2} \omega_z$}). 
However, the axial velocity produced by the slope effect is phase shifted by $\pi/2$ with $\omega_z$, and so does not allow the production of mean helicity. 
On the contrary, axial velocity produced by Ekman pumping is in phase with the axial vorticity and 
a dynamo mechanism based on the Ekman pumping associated to an azimuthal necklace of axial vortices is plausible \citep{Bus75}.
In a numerical experiment at small Ekman numbers ($E=\mathcal{O}(10^{-8})$), SC06 artificially remove the Ekman pumping and observe dynamo action with nearly the same threshold
showing that the Ekman pumping is unimportant in their dynamo mechanism. 
The crucial importance of the time dependence of the flow and the negligible contribution of the Ekman pumping lead 
SC06 to consider the involvement of the Rossby waves in the dynamo process.  
They conjecture that the propagation of the Rossby waves yields a proper phase shift between
 the non-axisymmetric magnetic field and velocity field in order to produce the axisymmetric poloidal magnetic field.

\citet{Ava09} have calculated the $\alpha$ tensor, describing the generation of a large scale magnetic field by correlation of small 
scale velocity and magnetic field, with a flow geometry corresponding to Rossby waves. In the absence of Ekman pumping, they 
show that the diagonal components of the $\alpha$ tensor, which are the relevant coefficients for the $\alpha$ effect, are non-zero 
if and only if the flow pattern is drifting relative to the mean flow. 

\citet{Til08} explains that the time dependence of a velocity field can lead to dynamo action even 
when any particular snapshot of the velocity field cannot because the linear operator associated with the induction equation is non-normal.
In particular, he shows that the simple time dependence of a propagating wave is enough for dynamo action.
Several numerical studies report the importance of the time dependence of the velocity field, mainly of
oscillating nature \citep{Reu09,Gub08}. 

The idea that the propagation of Rossby waves may maintain a dynamo action is very appealing as their
presence is ubiquitous in rotating fluid dynamics.
A system in which no wave propagation occurs, and which is unable to produce $u_z$ by another mechanism, such as buoyancy, will rely on Ekman pumping to create axial velocity with the proper phase shift. However, in the limit of small Ekman number, the Ekman pumping vanishes and the dynamo threshold should become infinitely high. 
The dynamo mechanism relying on the propagation of Rossby waves is robust in the limit of small Ekman number as the 
presence of these waves does not rely on the action of viscosity. 

Due to the close resemblance of the flow (zonal motions and propagating Rossby wave) 
in our 3D numerical model and the kinematic quasi-geostrophic model of SC06, 
we now try to establish if the dynamo mechanism evoked in SC06 is at work in our 3D model.
To formalize their idea, let us first consider a simple theoretical model. 
The velocity field is composed by a zonal flow, $\overline{U}(s) \vect{e_{\phi}}$, and the small scale 
velocity of a Rossby wave $\vect{u}^m$ with 
\begin{eqnarray}
\vect{u}^m(s,\phi,z,t)=(u_s^m (s,z)\vect{e_s}+u_{\phi}^m (s,z)\vect{e_{\phi}} + u_z^m (s,z)\vect{e_z}) 
			e^{i (m\phi-\omega t)}
\end{eqnarray}
where $u_s^m$, $u_{\phi}^m$ and $u_z^m$ are complex and $\omega$ is the frequency of the wave. 
The magnetic field is composed of an axisymmetric magnetic field $\overline{\vect{B}}$,
and a magnetic field perturbation induced at the scale of the Rossby wave $\vect{b}^m$ with 
\begin{eqnarray}
\vect{b}^m(s,\phi,z,t)=(b_s^m (s,z)\vect{e_s}+b_{\phi}^m (s,z)\vect{e_{\phi}} + b_z^m(s,z) \vect{e_z}) 
			e^{i (m\phi-\omega t) +\lambda t}
\end{eqnarray}
where $b_s^m$, $b_{\phi}^m$ and $b_z^m$ are complex and $\lambda$ is the growth rate of the magnetic field.
The equations for the evolution of the poloidal components of $\overline{\vect{B}}$ in cylindrical coordinates 
$\overline{B_s}$ and $\overline{B_z}$ are
\begin{eqnarray}
 \frac{\partial \overline{B_s}}{\partial t} &=& 
	- \frac{\partial}{\partial z}\left( \overline{u_z^m b_s^m - u_s^m b_z^m} \right) 
	+ \eta \left( \n^2 \overline{B_s} - \frac{\overline{B_s}}{s^2} \right),
\label{eq:B_s}
\\
 \frac{\partial \overline{B_z}}{\partial t} &=& 
	\frac{1}{s}\frac{\partial}{\partial s}s \left( \overline{u_z^m b_s^m - u_s^m b_z^m} \right) 
	+ \eta \n^2 \overline{B_z},
\label{eq:B_z}
\end{eqnarray}
where the overbar denotes an azimuthal average. 
It is immediately apparent that if $u_s^m$ ($u_z^m$) is out of phase by $\pi/2$ with $b_z^m$ (resp. $b_s^m$), then $\overline{B_{s}}$ and $\overline{B_{z}}$ will be decaying in time. 
If we suppose that $\overline{B_{\phi}}\gg \overline{B_s}, \overline{B_z}$ the equations 
for $b_s^m$ and $b_z^m$ are 
\begin{eqnarray}
 (\lambda - i c \frac{m}{s}) b_s^m &=& 
	\frac{i m}{s} u_s^m \overline{B_{\phi}}
	+ \eta \left(\n^2 b_s^m - \frac{2}{s^2}\frac{\partial b_{\phi}^m}{\partial \phi} - \frac{b_s^m}{s^2} \right),
\label{eq:b_s_m}
\\ (\lambda - i c \frac{m}{s}) b_z^m &=& 
	\frac{i m}{s} u_z^m \overline{B_{\phi}}
	+ \eta \n^2 b_z^m.
\label{eq:b_z_m}
\end{eqnarray}
where $c=(\omega/(m/s) -\overline{U})$ is the phase speed of the wave relative to the mean flow $\overline{U}$.
In the case of marginal stability ($\lambda=0$), if we neglect the magnetic diffusivity $\eta$ then we obtain that $b_s^m$ ($b_z^m$) is in phase with $u_s^m$ ($u_z^m$ resp.).
Moreover if the axial velocity is mainly due to the slope effect then $u_z^m=z\beta u_s^m$ and so according to
the equations~(\ref{eq:b_s_m})-(\ref{eq:b_z_m}) $b_z^m \approx z \beta b_s^m$. This implies that the first term of the right hand side of equations~(\ref{eq:B_s})-(\ref{eq:B_z})
is almost zero and thus $\overline{B_s}$ and  $\overline{B_z}$ are decaying. Consequently magnetic diffusivity at the scale of $b_s^m$ and $b_z^m$ must play
a role in the generation of the axisymmetric poloidal magnetic field by introducing a short phase lag between the velocity and magnetic modes.
This phase lag depends on the spatial structures of $b_s^m$ and $b_z^m$, and hence the terms
$\overline{u_s^m b_z^m}$ and $\overline{u_z^m b_s^m}$ do not cancel out. Note that the importance of magnetic diffusivity 
is well established in the $\alpha$ effect \citep{Rob07}.
On the other hand, if the wave is not propagating, $c=0$, then
\begin{eqnarray}
	- \frac{i m}{s} u_s^m \overline{B_{\phi}} &=&
	 \eta \left(\n^2 b_s^m - \frac{2}{s^2}\frac{\partial b_{\phi}^m}{\partial \phi} - \frac{b_s^m}{s^2} \right),
\label{eq:b_s_m_c0}
\\  
	- \frac{i m}{s} u_z^m \overline{B_{\phi}} &=&
	 \eta \n^2 b_z^m.
\label{eq:b_z_m_c0}
\end{eqnarray}
In this case the magnetic field perturbations $b_s^m$ and $b_z^m$ are out of phase with $u_s^m$ and $u_z^m$ (as 
$u_s^m$ and $u_z^m$ are correlated by the slope effect) and so the averaged products $\overline{u_z^m b_s^m}$ and $\overline{u_s^m b_z^m}$ are zero. 
We can conclude that in order for this simple model to work as a mean-field dynamo (i) the wave must propagate 
and (ii) the magnetic diffusivity must act on the magnetic field generated at the scale of the waves.
As the Rossby wave propagates, the location of 
the induction of the magnetic field perturbation is forced to drift with the same rate, but with a phase-shift.
The phase-shift between the magnetic field perturbation and the Rossby wave depends on both the phase speed $c$ 
and the magnetic diffusivity $\eta$. The argument above implies that this phase-shift is essential for the dynamo mechanism.

Using any particular snapshot of the velocity field for time stepping the magnetic induction in our 
numerical simulations with models J or N leads to the decay of the magnetic field. 
The failure of dynamo in the kinematic numerical experiment with both models is readily explained by our simple theoretical model.

\begin{figure}
 \centering
  \subfigure[Model J]{\label{fig:bsus_ProfJ}
   \includegraphics[clip=true,width=0.55\textwidth]{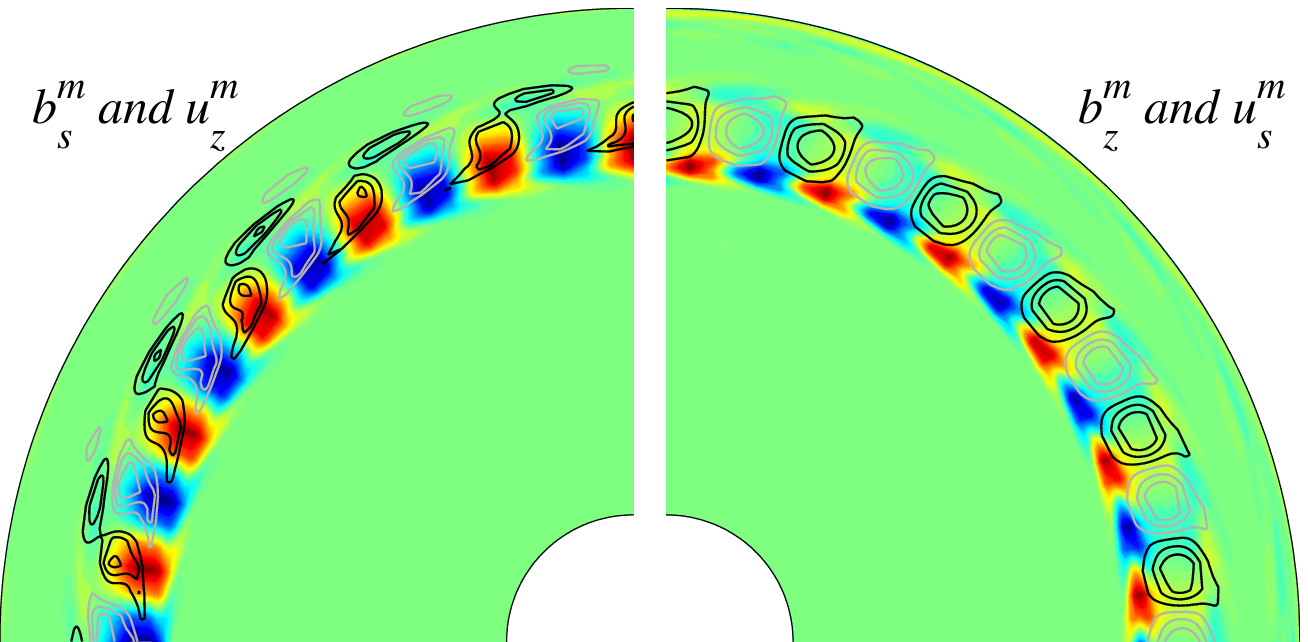}}
   \hspace{0.5cm}
  \subfigure[Model N]{\label{fig:bsus_ProfN}
  \includegraphics[clip=true,width=0.35\textwidth]{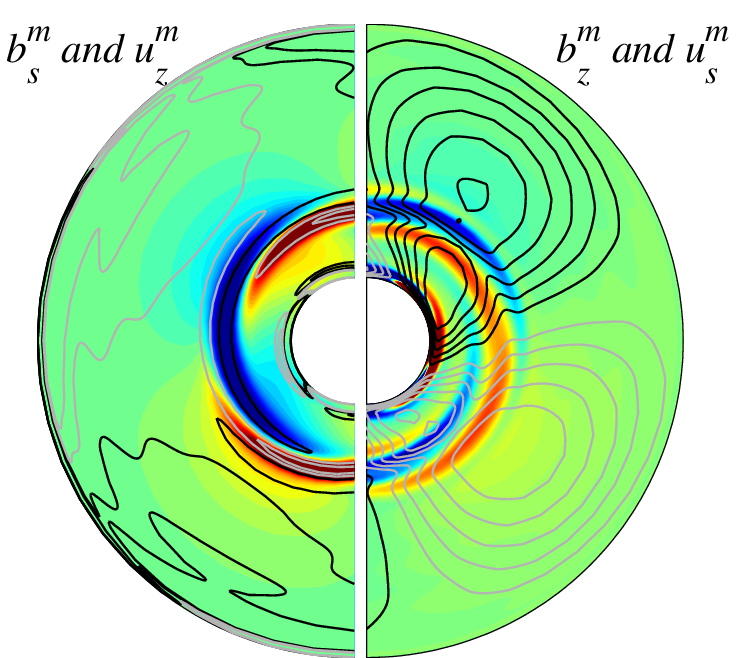}}
  \caption{Non-axisymmetric magnetic field (coloured) and non-axisymmetric velocity (black lines: positive, and gray lines: negative) 
  in a plane a few degree of latitude above the equatorial plane (northern hemisphere). 
  Same parameters than figures~\ref{fig:B_ProfJ} and~\ref{fig:B_ProfN}.}
\label{fig:mecanisme_alpha}
\end{figure}

In figure~\ref{fig:mecanisme_alpha}, we plot the non-axisymmetric components of the velocity, $u_z^m$ and $u_s^m$ and magnetic field, 
$b_s^m$ and $b_z^m$ obtained in the numerical simulations for model J and model N
in a plane located just above the equatorial plane 
($b_s^m$ and $b_z^m$ are zero in the equatorial plane by dipolar symmetry in model J).
The correlation of $u_z^m$ with $u_s^m$ confirms that 
$u_z^m$ is mainly produced by the slope effect for both models. 
For model J (figure~\ref{fig:bsus_ProfJ}), we observe that $u_z^m$ and $b_s^m$ are in phase so $\overline{u_z^m b_s^m}$ has a significant amplitude.
However, $b_z^m$ is out of phase with $u_s^m$, which means 
that $\overline{u_z^m b_s^m} \gg \overline{u_s^m b_z^m}$.
This may be an effect of the magnetic diffusivity as $b_s^{m}$ and $b_z^{m}$ have 
different spatial structures, or 
due to radial derivatives of $\overline{B_s}$ and $\overline{B_z}$ that we neglect in equation~(\ref{eq:b_z_m}).
Consequently $\overline{u_z^m b_s^m}$ mainly contributes to the generation of strong $\overline{B_s}$ and $\overline{B_z}$. 

For model N (figure~\ref{fig:bsus_ProfN})
strong positive (negative) crescent-shaped patches of $b_s^m$ and $b_z^m$ 
are visible in the cyclonic (resp. anticyclonic) vortices, out of phase by $\pi/2$ with $u_s^m$ and $u_z^m$. 
Consequently these crescent-shaped structures of $b_s^m$ and $b_z^m$ do not contribute to the terms $\overline{u_z^m b_s^m}$ and $\overline{u_s^m b_z^m}$. 
The presence of these maxima of $b_s^m$ and $b_z^m$ are not explained by the theoretical model (equations~(\ref{eq:b_s_m}) and (\ref{eq:b_z_m}))
likely because of the neglect of the axial and radial derivatives of $\overline{B_s}$ and $\overline{B_z}$, which are important in this region (see figure~\ref{fig:Baxi_ProfN}).
Round-shaped lobes of $b_s^m$ and $b_z^m$ of weaker amplitude (located in the middle of the gap) are observed in phase with $u_s^m$ and $u_z^m$. 
Consequently these round-shaped structures of $b_s^m$ and $b_z^m$ contribute to the terms $\overline{u_z^m b_s^m}$ and $\overline{u_s^m b_z^m}$. 
Unlike model J (where $\overline{u_z^m b_s^m} \gg \overline{u_s^m b_z^m}$), $\overline{u_z^m b_s^m} \sim \overline{u_s^m b_z^m}$
so only a weak axisymmetric multipolar magnetic field is maintained in this case.
At the surface of the planet this axisymmetric field is the dominant 
component but in comparison with the strongly axisymmetric dipolar field produced in model J, the field
is of small amplitude: the amplitude of the axisymmetric radial field is about $10^{-3}\sqrt{\rho \mu_0}U_0$ for model J 
at $Rm=1.17Rm_c$ ($Ro=1.17Ro_c$ and $Pm\approx Pm_c$) (figure~\ref{fig:spec_rs_J}) while it is only $10^{-5}\sqrt{\rho \mu_0}U_0$ 
for model N at $Rm_c=2.4Rm_c$ ($Ro=1.20 Ro_c$ and $Pm=2 Pm_c$) (figure~\ref{fig:spec_rs_N}).

The main difference between the Rossby waves in models J and N is their size. 
The phase speed of the Rossby wave, $c\approx \Omega \beta/(m/s)^2$,
is about 100 times larger for a $m=2$ wave than a $m=22$ wave, for a fixed radius $s$ and rotation rate $\Omega$.
On the other hand, the magnetic diffusion acts more rapidly on small scale structures.
The typical propagation timescale for a Rossby wave of size $d$ is $\tau_{rw}=1/(\Omega \beta d)$ assuming
that the radial and azimuthal lengthscales of the wave are similar. The magnetic diffusion timescale
at the scale of the vortex $d$ is $\tau_{\eta}=d^2/\eta$. The ratio of the two timescales is
\begin{eqnarray}
 \frac{\tau_{\eta}}{\tau_{rw}} = \frac{d^3 \Omega \beta}{\eta} .
\end{eqnarray}
The dependence to the third power of the size, $d\propto 1/m$, shows that the magnetic diffusion timescale relative 
to the propagation timescale is about three orders of magnitude smaller for an $m=22$ mode than an $m=2$ mode for the same parameter values.
For the simulation presented for model N, the ratio $\tau_{\eta}/\tau_{rw}$ is about $10^5$.
For model J the ratio $\tau_{\eta}/\tau_{rw}$ is about $500$ 
so the propagation of the Rossby wave is still much more rapid than the magnetic diffusion.
For both models, we found that the values of the small scale magnetic field in phase with the velocity
is of the same order of magnitude. The velocity field of the vortices is also about the
same order of magnitude for the two models. 
The difference between the two models is that, in model N, the magnetic diffusion acts too slowly on the $m=2$ 
magnetic structures compared to the wave propagation to
produce a significant enough phase lag between $b_s^m$ ($b_z^m$) and $u_z^m$ ($u_s^m$).
Consequently, the term $\overline{u_s^m b_z^m}-\overline{u_z^m b_s^m}$ is weak
and leads to little generation of axisymmetric poloidal magnetic field.

\begin{figure}
 \centering
  \subfigure[Model J]{\label{fig:omega_ProfJ}
  \includegraphics[clip=true,width=0.2\textwidth]{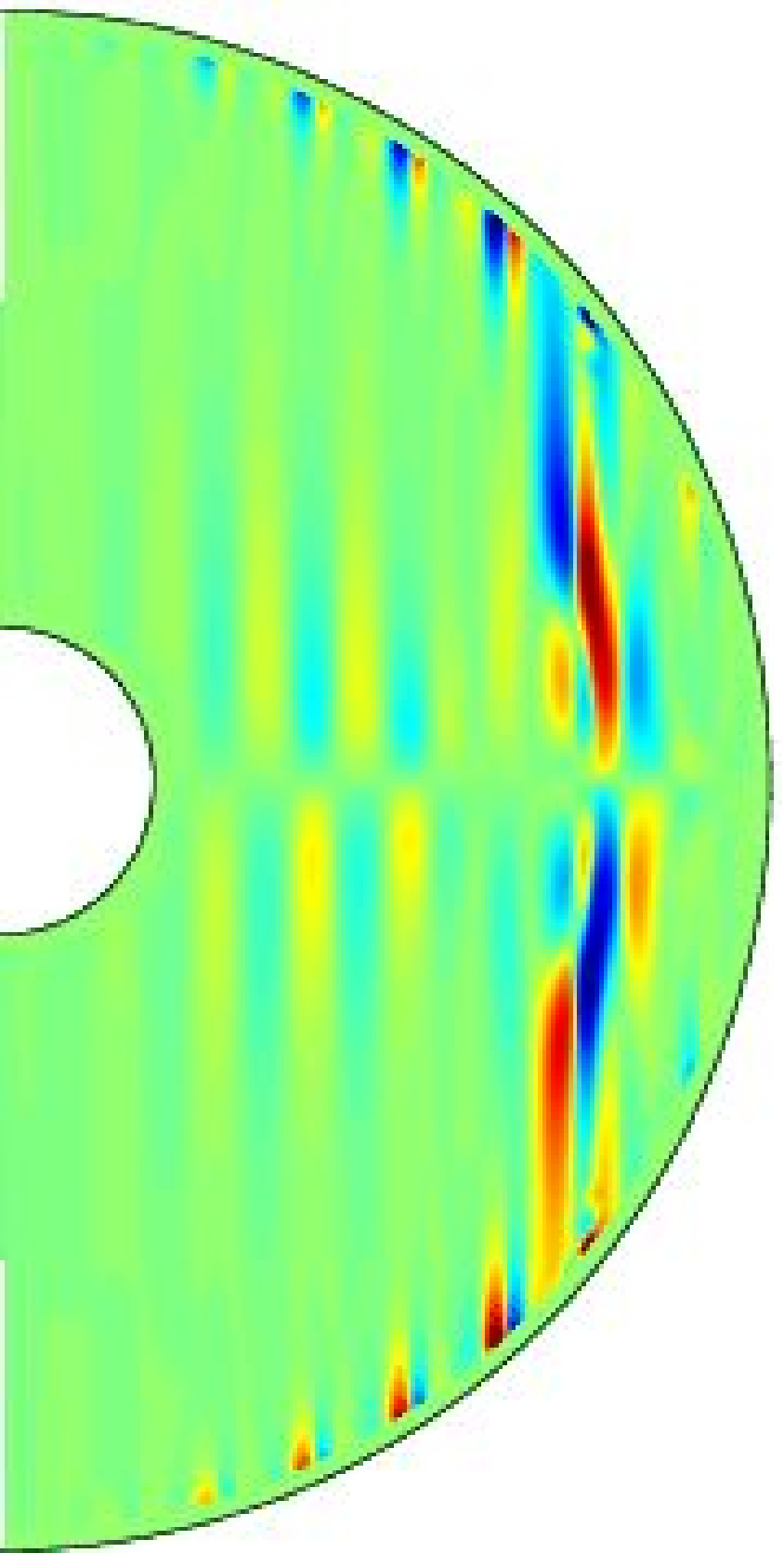}} 
  \hspace{0.5cm}
  \subfigure[Model N]{\label{fig:omega_ProfN}
  \includegraphics[clip=true,width=0.2\textwidth]{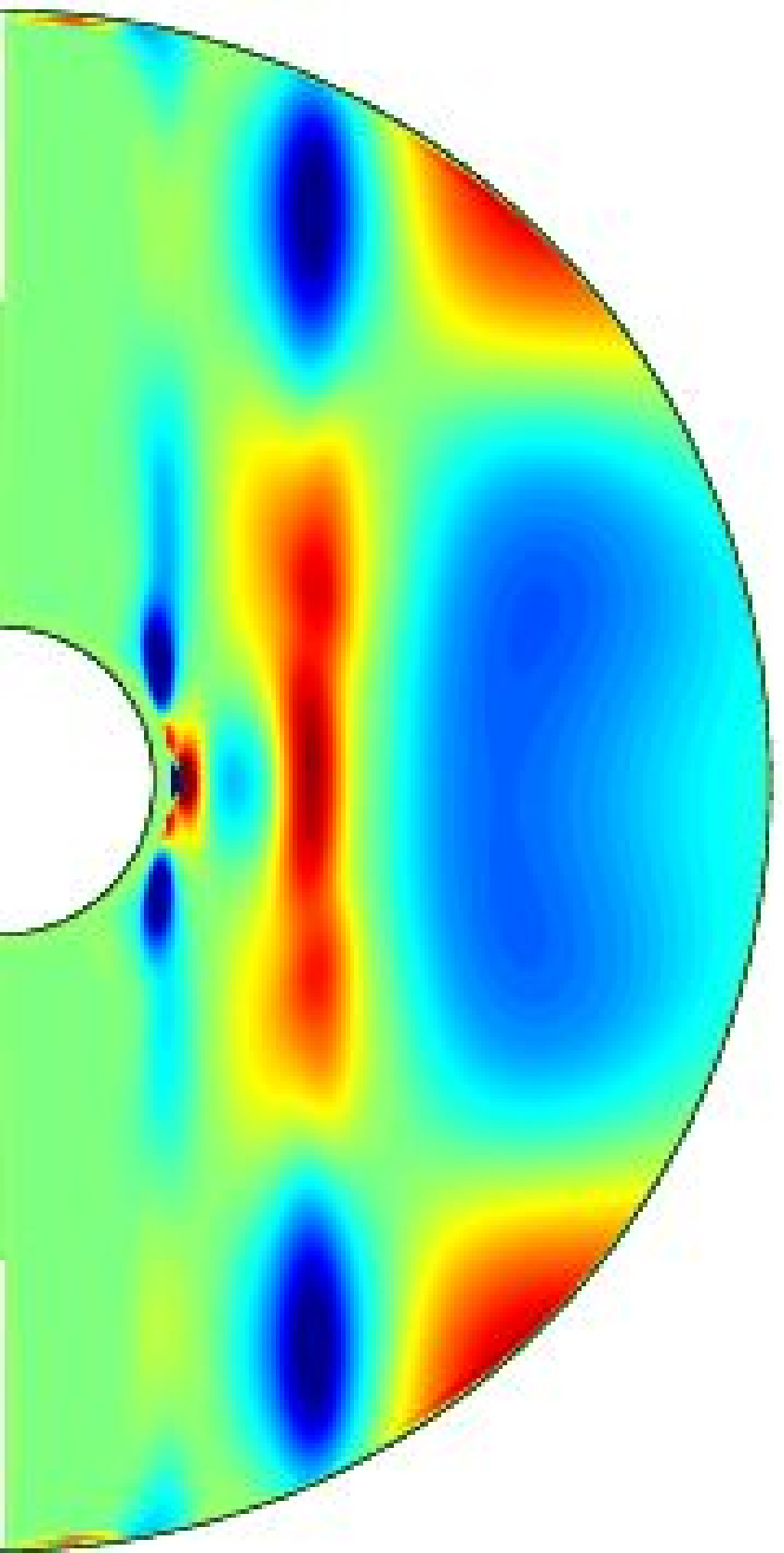}}
  \caption{$\omega$ effect in the meridional plane in the bulk (outside the Ekman layers) 
  (blue: negative and red: positive).
Same parameters than figures~\ref{fig:B_ProfJ} and~\ref{fig:B_ProfN}.}
 \label{fig:mecanisme_omega}
\end{figure}

The last stage of the dynamo mechanism is the generation of the axisymmetric toroidal field. 
It can either be produced from the correlation of small scale velocity and magnetic field (as an $\alpha$ effect)
 or an $\omega$ effect, that is the shearing of the axisymmetric poloidal magnetic field by the mean zonal flow $\overline{U}$. 
SC06 find that the $\omega$ effect from the Stewartson layer is dominant in their numerical model.
The zonal shear produced in the Stewartson layer is stronger than the shear we obtained with the profiles J and N, 
so it is not clear that the $\omega$ effect is important in our model \textit{prima facie}.
In $\alpha^2$ dynamos, both toroidal and poloidal components are typically of similar magnitudes \citep{Ols99}.
Here, the strong toroidal magnetic field suggests that the $\omega$ effect is more important. 
To confirm this, we plot in figure~\ref{fig:mecanisme_omega} the term responsible for the $\omega$ effect in the azimuthal 
component of the magnetic induction equation \citep{Gub87}, $r \overline{B_r} \partial_r (r^{-1} \overline{U}) + r^{-1} \sin \theta \overline{B_{\theta}} \partial_{\theta} (\sin \theta^{-1} \overline{U})$. For model J, as we expect, this
term is most significant in the region where the poloidal magnetic field lines are bent and misaligned with the zonal flow structure (see figure~\ref{fig:Baxi_ProfJ}).
The correlation of sign and location of the maxima of the $\omega$ effect in the bulk of the fluid with 
the axisymmetric azimuthal field indicates that it is mainly generated by the $\omega$ effect. 
Note that some $\omega$ effect is also present close to the outer boundary, where the poloidal magnetic field lines
converge and diverge locally due to induction by the Ekman pumping. However no particularly strong axisymmetric
azimuthal magnetic field is produced in this region (figure~\ref{fig:Baxi_ProfJ}) so this small scale field
diffuses probably very rapidly.
For model N the outer part of the jet ($s>0.5$) is retrograde and creates a negative $\omega$ effect whose 
sign and location correlate with the axisymmetric azimuthal field, implying that the main dynamo process in the outer 
region is indeed the $\omega$ effect. However, $\overline{B_{\phi}}$ and the $\omega$ effect are anti-correlated in the 
inner region ($s<0.5$) so another dynamo process such as a correlation of small scale velocity and magnetic field must be at work there.

We have not yet addressed the question of the selection of the axial dipolar symmetry or the
axial quadrupolar symmetry.
In kinematic dynamo calculations, \citet{Gub00} show that minor changes in the flow can select very different
eigenvectors. For a self-consistent system the selection rules are thus very subtle.
As discussed in section~\ref{sec:B_profN}, \citet{Aub04} found that axial quadrupolar symmetry is incompatible 
with the vertical structures of cyclones and anticyclones in convectively-driven dynamos, and so these solutions are 
unstable for strong convective flows. In our simulations of model N, this
conclusion suggests that the axial quadrupolar symmetry would be unstable for larger magnetic
Reynolds numbers, and an axial dipolar field would be preferred. The selection of a given symmetry does not modify 
our argument that the wavenumber of the Rossby mode determines the amplitude of the axisymmetric magnetic field since no
particular latitudinal symmetry is assumed.

In this study, it appears that the dynamo mechanism relies on a subtle balance between the Rossby wave propagation 
and the magnetic diffusion and therefore is closely related to the size of the Rossby waves.
The dynamo field produced with this mechanism requires high magnetic Reynolds numbers ($Rm_c\approx20,000$ for
model J and $Rm_c\approx 4000$ for model N). However, in the limit of small Ekman number, this dynamo mechanism is 
expected to keep a finite value of the critical magnetic Reynolds number 
\citep{Sch06}, whereas for dynamos that rely on Ekman pumping the critical magnetic Reynolds number becomes infinitely high. 

\section{Summary and discussion}
\label{sec:Summary}
We have numerically studied the dynamics of zonal flows driven by differential rotation imposed at the top of a conducting layer and how they sustain a magnetic field. 

\subsection{Hydrodynamical instability}
In our hydrodynamical simulations, we found that the destabilisation of the zonal flow takes the form of a global (large radial extension) Rossby mode, even though the instability threshold is governed by a local criterion. The wavenumber depends on the width of the jets, and is independent of the viscosity and rotation rate provided that 
the former is sufficiently small. 
In the supercritical regime, 
several Rossby waves appear and saturate the amplitude of the zonal flow in the bulk of the fluid. They produce a widening of the jets and a strong damping of their amplitude, even for relatively small supercritical forcing ($Ro=2.94Ro_c$).

\subsection{Constraints on the dynamo mechanism}
In the limit of small Ekman number, we find that the Rossby wave appears for $Ro_c\approx0.001$ for a Jupiter-like zonal 
wind profile (model J) and $Ro_c\approx0.02$ for a Neptune-like profile (model N). In our numerical calculations, non-axisymmetric 
motions are necessary for dynamo action to occur.
As the viscosity is large in the numerical simulations compared to the planetary values, the Reynolds number is much smaller in the simulations.
To reach a sufficiently high magnetic Reynolds number, the magnetic Prandtl number is of order $1$, much larger than the expected planetary values. 
The critical magnetic Reynolds number $Rm_c$ is about $20,000$ for model J and $4000$ for model N.
To make this dynamo mechanism work, two constraints must be satisfied: (i) $Ro>Ro_c$ and (ii) $Rm>Rm_c$. 
Equivalently this gives constraints on the amplitude of the zonal motions at the top of the conducting region, $U_0>Ro_c\Omega r_o$, and
on the electrical conductivity within the conducting region, $\sigma>Rm_c/(U_0 r_o \mu_0)$.

The extrapolation of the constraint (i) to the giant planets is straightforward as the hydrodynamical instability threshold is independent of the 
Ekman number, which is of order $10^{-15}-10^{-16}$ for Jupiter \citep{Gui04} and Neptune \citep{Ste83}. 
For Jupiter ($r_o \approx 56,000$ km and $\Omega=1.8\times10^{-4}$s$^{-1}$), the equatorial velocity at the top of the conducting region, $U_0$, must be larger than $10$ m/s 
to have $Ro>Ro_c=0.001$. 
For Neptune ($r_o \approx 21,000$ km and $\Omega=1.08\times10^{-4}$s$^{-1}$), $U_0$ must be larger than $45$ m/s to have $Ro>Ro_c=0.02$.
For both cases, this constraint is quite strong as it only allows for a factor 10 decrease of the amplitude of the zonal wind between the surface of the planet and the top of the 
deep conducting region, independently of the location of the top of this region. 

The extrapolation of the constraint (ii) to the giant planets requires knowing how the critical Reynolds number scales with the Ekman number.
When varying the Ekman number from $10^{-6}$ down to $10^{-8}$, \citet{Sch06} found that $Rm_c$ remains constant (of the order of $10^4$ 
in their simulations, close to the values found in our study). Based on their results, we assume that $Rm_c$ is of the same order of magnitude 
when the Ekman number is close to the planetary values. 
For Jupiter, we obtain that the electrical conductivity should be larger than $30$S/m to have $Rm>Rm_c=20,000$ (using $U_0=10$ m/s). 
For Neptune, the electrical conductivity should be larger than $10$S/m to have $Rm>Rm_c=4000$ (using $U_0=45$ m/s). 
This constraint on the conductivity is less restrictive than the constraint on the amplitude of the zonal motions
and should be satisfied in the deep conducting layer of Jupiter \citep{Nel99} and Neptune \citep{Nel97}.

We conclude that the differential rotation imposed by the zonal winds at the top of the conducting regions is a plausible candidate
to drive the dynamo mechanism 
in the giant planets although a strong constraint on the amplitude of the zonal jet applies.
Given the assumptions used in our model, such as incompressibility, constant conductivity, unrealistically large viscosity
and viscous coupling between electrically insulating and conducting regions,
this conclusion remains tentative. 
However, the robust nature of Rossby waves in the asymptotic limit of small Ekman numbers makes this dynamo mechanism appealing 
for planetary physical conditions.

\subsection{Generation of the axisymmetric field and width of the jets}
With a simple theoretical model, we show that the production of the axisymmetric field depends on the propagation of the Rossby waves and 
on the magnetic diffusion acting at the scale of the vortices. This model is in agreement with our numerical results:
the magnetic diffusion rate of the $m=2$ magnetic structures induced by the Rossby waves in model N is nearly negligible 
compared to the propagation rate of the wave: as a result a weak axisymmetric poloidal magnetic field is generated;
the magnetic diffusion acting on the $m=22$ magnetic structures is not negligible compared to the propagation rate of the small size ($m=22$) Rossby wave of model J: 
a dominant axisymmetric poloidal magnetic field is therefore generated. 
Consequently, in this model, the width of the zonal jets has an important influence on the generation of the axisymmetric magnetic field by controlling the size of the Rossby waves. 
Our results suggest that the difference in the magnetic fields and the surface zonal winds may be related if a (hydrodynamic or magnetohydrodynamic) mechanism 
can transport angular momentum between the surface and the deep, electrically conducting region.

The critical magnetic Reynolds number of this dynamo mechanism is large.
However, in order to compare with other dynamos, a more significant number may be the critical local magnetic Reynolds number associated with magnetic induction by the Rossby wave velocity
$Rm_c^l=V_s d/\eta$ where $V_s$ is the typical non-axisymmetric radial velocity and $d$ is the lengthscale of the Rossby mode. 
For the dynamo obtained in model J ($Rm_c= 20,000$), $V_s\approx0.1U_0$ and $m=22$ so 
we find $Rm_c^l \approx 570$. 
For the dynamo obtained in model N ($Rm_c= 4000$), $V_s\approx0.01U_0$ and $m=2$ so 
$Rm_c^l \approx 130$.
Thus $Rm_c^l$ is roughly $2-10$ times larger than the magnetic Reynolds number needed for dynamo action with a convective forcing \citep{Chr06}.

\subsection{Magnetic field at the planets' surfaces}
In our numerical model, we obtain a peak at small azimuthal scale in the magnetic field spectrum correlated with the width of the hydrodynamically unstable zonal jets.
This is a testable prediction as the magnetic measurements of the forthcoming \textit{Juno} mission (arrival at Jupiter in 2016) are expected to be of extraordinary quality 
due to the absence of a crustal magnetic field on Jupiter. 

For model J, we obtain a secular variation of the dipole tilt of about $1^{\circ}$ in $1000$ rotation periods
or equivalently $0.001$ global magnetic diffusion time.
The dipole is strongly axisymmetric with a tilt that does not exceed $2^{\circ}$. 
On Jupiter, the dipole axis tilt measured with the \textit{Pioneer} and \textit{Voyager} data compared with the \textit{Galileo} measurements is larger (about $10^\circ$) and 
displays a secular variation of about $0.5^{\circ}$ in 20 years \citep{Rus01}.
The strong axisymmetry of the dipolar field of model J is in better agreement with the magnetic field of Saturn with a dipole tilt less than $1^{\circ}$\citep{Rus10}.

\subsection{Convective motions within the conducting region}
In this work we have not taken into account the convective motions within the deep conducting region. 
\citet{Wic02b} studied the linear stability of an imposed zonal flow in a spherical shell modeling the molecular hydrogen layer of Jupiter. 
They found that the critical Rossby number of the shear instability onset is almost independent of the Rayleigh number, which measures 
the strength of the convection. They concluded that the shear instability is only weakly modified by the presence of convection. 
On the other hand, they showed that the convection onset is strongly influenced by the presence of the zonal circulation, with the convection 
either enhanced or damped depending on the direction of the shear.
However, their study is linear, and so the results cannot be extrapolated beyond the weakly non-linear regime of convection. 
Whether or not our results apply in the presence of convection is a subject for future studies.
In the presence of convection (which produces strong zonal motions and Rossby waves), and even for a convectively-driven dynamo \citep[see for instance][]{Aub05,Gro01},
the mechanism described here may still impose a similar relationship between the 
magnetic field morphology and the zonal wind profile.

\section*{Acknowledgments}
Financial support was provided by the Programme National de Plan\'etologie of CNRS/INSU. C.G. was supported by a research studentship from Universit\'e Joseph-Fourier Grenoble and by the Center for Momentum Transport and Flow Organization sponsored by the US Department of Energy - Office of Fusion Energy Sciences. The computations presented in this article were performed at the Service Commun de Calcul Intensif de l'Observatoire de Grenoble (SCCI) and at the Centre Informatique National de l'Enseignement Sup\'erieur (CINES).
We thank Jonathan Aurnou, Toby Wood and the geodynamo group in Grenoble for useful discussions. 
The manuscript was substantially improved due to helpful suggestions by two anonymous referees.
This is a preprint of an article whose final and definitive form has been published in Icarus. 
Icarus is available online at: \href{http://www.journals.elsevier.com/icarus}{http://www.journals.elsevier.com/icarus}.

\appendix

\section{Linear code used to compute the hydrodynamical instability threshold}
\label{app:codeXSHELL}
In order to study the linear stability threshold at very low Ekman numbers, we designed a linear code derived from \citet{Gil11}.
This three-dimensional spherical code uses second order finite differences in radius and pseudo-spectral spherical harmonic expansion.
The linear perturbation $\mathbf{u}$ of the imposed background flow $\mathbf{U}$ is time-stepped from a random initial field with the following equation:
\begin{eqnarray}
\left( \frac{\partial}{\partial t} - \mathbf{\nabla}^2 \right) \mathbf{u} = 
-\left( \frac{2}{E} \mathbf{e_z} + \mathbf{\nabla} \times \mathbf{U} \right) \times \mathbf{u} 
+ \mathbf{U} \times \mathbf{\nabla} \times \mathbf{u} -\mathbf{\nabla} p,
\label{eq:codelin}
\end{eqnarray}
together with the continuity equation $\mathbf{\nabla} \cdot \mathbf{u} = 0$, which allows us to eliminate the pressure term by using a poloidal-toroidal decomposition.
The left hand side of equation~\ref{eq:codelin} is treated with a semi-implicit Crank-Nicolson scheme, whereas the right hand side is treated as an explicit Adams-Bashforth term.
Thanks to the cylindrical symmetry of the base flow $\mathbf{U}$, all azimuthal modes $m$ of the perturbation $\mathbf{u}$ are independent, and we can compute them separately.
The coupling with the background flow and the Coriolis force are handled in physical space, but a very fast implementation of the spherical harmonic transform (SHTns library) makes the code quite efficient.

In order to determine the stability threshold at $E=10^{-7}$, we used $350$ points in the radial direction, and spherical harmonics truncated at \mbox{$l_{max} = 300$}.
We use no-slip boundary conditions.
 
\bibliographystyle{model2-names.bst}
\bibliography{ref}
 
\end{document}